\documentclass[fleqn,usenatbib]{mnras}

\usepackage{newtxtext,newtxmath}

\usepackage[T1]{fontenc}

\DeclareRobustCommand{\VAN}[3]{#2}
\let\VANthebibliography\thebibliography
\def\thebibliography{\DeclareRobustCommand{\VAN}[3]{##3}\VANthebibliography}

\usepackage{graphicx}	
\usepackage[capitalise]{cleveref}
\crefname{equation}{equation}{equations}
\usepackage{makecell}
\usepackage{multirow}
\usepackage{amsmath}	
\usepackage[normalem]{ulem}

\title[PAU Survey: D4000 measurement]{The PAU Survey: Measurements of the 4000 \AA\ spectral break with narrow-band photometry}

\author[P. Renard et al.]{
Pablo Renard$^{1,2,3}$\thanks{E-mail: p.renard.guiral@gmail.com},
Małgorzata Siudek$^{4,5}$,
Martin B. Eriksen$^{4}$,
Laura Cabayol$^{4}$,
\newauthor
Zheng Cai$^{1}$,
Jorge Carretero$^{4,6}$,
Ricard Casas$^{2,3}$,
Francisco J. Castander$^{2,3}$,
\newauthor
Enrique Fernandez$^{4}$,
Juan García-Bellido$^{7}$,
Enrique Gaztanaga$^{2,3}$,
Henk Hoekstra$^{8}$,
\newauthor
Benjamin Joachimi$^{9}$,
Ramon Miquel$^{4,10}$,
David Navarro-Girones$^{2,3}$,
Crist\'obal Padilla$^{4}$,
\newauthor
Eusebio Sanchez$^{11}$,
Santiago Serrano$^{2,3}$,
Pau Tallada-Cresp\'{i}$^{11,6}$,
Juan De Vicente$^{11}$,
\newauthor
Anna Wittje$^{12}$,
and Angus H. Wright$^{12}$
\\
$^{1}$Department of Astronomy, Tsinghua University, Beijing 100084, China\\
$^{2}$Institute of Space Sciences (ICE, CSIC), Carrer de Can Magrans s/n, E-08193 Bellaterra (Barcelona), Spain\\
$^{3}$Institut d'Estudis Espacials de Catalunya (IEEC), E-08034 Barcelona, Spain\\
$^{4}$Institut de Física d’Altes Energies (IFAE), The Barcelona Institute of Science and Technology, 08193 Bellaterra (Barcelona), Spain\\
$^{5}$National Centre for Nuclear Research, ul. Pasteura 7, 02-093 Warsaw, Poland\\
$^{6}$Port d'Informaci\'{o} Cient\'{i}fica (PIC), Campus UAB, C. Albareda s/n, 08193 Bellaterra (Barcelona), Spain\\
$^{7}$Instituto de Física Teórica UAM/CSIC, Universidad Autónoma de Madrid, 28049 Madrid, Spain\\
$^{8}$Leiden Observatory, Leiden University, Niels Bohrweg 2, 2333 CA, Leiden, The Netherlands\\
$^{9}$Department of Physics and Astronomy, University College London, Gower Street, London WC1E 6BT, UK\\
$^{10}$Institució Catalana de Recerca i Estudis Avançats (ICREA), 08010 Barcelona, Spain\\
$^{11}$Centro de Investigaciones Energéticas, Medioambientales y Tecnológicas (CIEMAT), 28040 Madrid, Spain\\
$^{12}$Ruhr University Bochum, Faculty of Physics and Astronomy, Astronomical Institute (AIRUB), German Centre for Cosmological Lensing,\\
44780 Bochum, Germany\\
}

\date{Accepted XXX. Received YYY; in original form ZZZ}

\pubyear{2021}

\begin{document}
\label{firstpage}
\pagerange{\pageref{firstpage}--\pageref{lastpage}}
\maketitle

\begin{abstract}
The D4000 spectral break index is one of the most important features in the visible spectrum, as it is a proxy for stellar ages and is also used in galaxy classification. However, its direct measurement has always been reserved to spectroscopy. Here, we present a general method to directly measure the D4000 with narrow-band (NB) photometry; it has been validated using realistic simulations, and then evaluated with PAUS NBs, cross-matched with VIPERS spectra ($i_{\rm AB} < 22.5$, $0.562 < z < 0.967$). We also reconstruct the D4000 with the SED-fitting code CIGALE; the use of PAUS NBs instead of broad bands significantly improves the SED fitting results. For D4000$_{\rm n}$, the direct measurement has $\rm \langle SNR \rangle \sim 4$, but we find that for $i_{\rm AB}<21$ all direct D4000 measurements have $\rm SNR>3$. The CIGALE D4000$_{\rm n}$ has $\rm \langle SNR \rangle \sim 20$, but underestimates the error by $>$50\%. Furthermore, the direct method recreates well the D4000-SFR relation, as well as the D4000-mass relation for blue galaxies (for red galaxies, selection effects impact the results). On the other hand, CIGALE accurately classifies galaxies into red and blue populations. We conclude that the direct measurement of D4000 with narrow-band photometry is a promising tool to determine average properties of galaxy samples, with results compatible with spectroscopy.
\end{abstract}




\section{Introduction}

In a broad sense, galaxy observations are divided in two different methodologies: spectroscopy and photometry. Spectroscopic surveys (e.g., DEEP2 \citep{Davis2003}, GAMA \citep{Baldry2010} eBOSS \citep{Dawson2016a}, VIPERS \citep{Scodeggio2018}, VANDELS \citep{Pentericci2018}) provide enough spectral resolution to clearly distinguish emission lines, breaks, etc. from the spectral continuum, as well as to determine with high precision the redshifts of the observed galaxies, to the point of redshift errors being considered negligible for most applications. Nevertheless, spectroscopic observations require large integration times for each object, as well as previous target selection.

On the other hand, most photometric surveys observe with broad-band filters (e.g. KiDS \citep{DeJong2013}, DES \citep{Abbott2018}, DESI Legacy Imaging Surveys \citep{Dey2019}) which can observe entire sky fields with far less telescope time, thanks to the large full width at half maximum (FWHM) of the filters used (roughly of the order of $\sim 1000$ \AA). This filter width allows to obtain high signal-to-noise ratio (SNR) with relatively low integration times, but also prevents the observation of any spectral features other than the continuum, given the low spectral resolution.

A middle ground between spectroscopic and broad-band photometric surveys are narrow-band photometric surveys, with photometric filters of the order of $\rm FWHM \sim 100$ \AA. These offer much larger spectral resolution and higher redshift precision than their broad-band counterparts, without increasing observation time as much as spectroscopy. They also keep other advantages of photometric surveys: no specific target selection is required, and the reduction of their images is still significantly less complex than that of spectra. Most existing narrow-band surveys so far observed small areas of the order of few deg$^2$, covering the visible spectrum with medium-width filters of $\rm FWHM > 200$ \AA, such as COMBO-17 \citep{Wolf2003} or ALHAMBRA \citep{Moles2008}. Also, surveys specifically designed for the observation of certain emission lines with very few filters have also been carried out, like SILVERRUSH \citep{Ouchi2018} for the Ly$\alpha$ line.

Narrow-band surveys covering significant angular areas, sampling the whole visible spectrum with filters of $\rm FWHM \sim100$ \AA, have been proposed during the past decade, and are currently starting to reveal its potential. One example of these surveys is the ongoing Physics of the Accelerating Universe Survey \citep[PAUS,][]{Benitez2009}, whose data is employed in this work. PAUS has been primarily designed to determine high-precision photometric redshifts \citep{Marti2014, Eriksen2019}, mostly allowing for precise cross-correlation of lensing and redshift distortion probes \citep{Gaztanaga2012}. Other examples of narrow-band surveys with larger angular coverage are J-PAS \citep{Benitez2014}, J-PLUS \citep{Cenarro2019} and S-PLUS \citep{DeOliveira2019}. These more general-purpose narrow-band surveys show potential to resolve specific spectral features, with closer approximation to spectroscopy \citep[e.g.,][]{Stothert2018, Martinez-Solaeche2020}.

The observation of specific spectral features besides the spectral continuum allows us to determine key galaxy properties such as star formation rate (SFR), stellar age and metallicity \citep[e.g.,][]{Worthey1994, Kauffmann2003a}; these spectral features are the results of different physical processes, and thus are a proxy to study the formation and evolution of galaxies. Generally, emission lines provide information about the ionised interstellar medium \citep[ISM;][]{Kewley2019}, while absorption lines inform of the properties of the stellar population \citep{Maraston2009}. For example, emission lines are commonly used to select active galactic nuclei (AGN) based on the Baldwin-Phillips-Terlevich (BPT) diagram \citep{Baldwin1981, Kauffmann2003, Kewley2013}, since they allow to discern between ionisation in the intergalactic medium (IGM) due to a soft UV source (e.g. star formation) or hard UV (e.g. AGN emission). On the other hand, absorption Lick indices are proxies for inferring stellar ages and metallicities \citep{Worthey1994, Thomas2003}.

The direct measurement of spectral features is generally reserved to spectroscopic surveys. However, galaxy properties can also be estimated with other methods, where the entirety of the spectrum is fit to stellar population models (full spectrum fitting). These fits are often convolved with the line-of-sight velocity distribution, which makes it especially useful to study galaxy kinematics. Some examples of full spectrum fitting codes are \textsc{STECKMAP} \citep{Ocvirk2006}, \textsc{ULySS} \citep{Koleva2009} and \textsc{pPFX} \citep{Cappellari2017}. Perhaps the most prominent one is spectral energy distribution (SED) fitting, in which a linear combination of spectral templates associated with different physical systems (stellar population models, dust emission/attenuation, AGNs, etc.) are fit to the observed spectral energy distribution, and the galaxy properties are inferred from the fit. This is a method widely applied to broad-band photometry, and several codes exist in the literature for this purpose, such as \textsc{LePHARE} \citep{Ilbert2006}, \textsc{CIGALE} \citep{Noll2009, Boquien2019} or \textsc{ProSpect} \citep{Robotham2020}; moreover, the application of SED-fitting codes for narrow-band photometric surveys has already been evaluated in \citet{Delgado2021} for the mini JPAS data release.

One spectral feature of special interest for this work are \textit{spectral breaks}: combinations of absorption lines in broader features that damp the continuum emission at given wavelength ranges. The most important of these features in the optical part of galaxy spectra is the 4000 \AA\ break, generated around that wavelength range by the absorption of several ionised metallic elements, as well as the latest lines in the Balmer series. The strength of this spectral break is quantified by the ratio between the continuum fluxes before and after the break itself; this quantity is known as the D4000 spectral index, or just \textit{D4000}. Two definitions for the D4000 with slightly different wavelength ranges exist in the literature: the original D4000 wide definition \citep{Bruzual1983, Hamilton1985}, and another defined in a narrower wavelength span \citep{Balogh1999}, which is the current standard. Henceforth, we will refer to these definitions as D4000$_{\rm w}$ and D4000$_{\rm n}$ respectively.

While D4000$_{\rm n}$ is generally preferred, mostly because it is far less sensitive to reddening effects due to the smaller continuum regions, both D4000 definitions have been extensively applied to study galaxy properties (e.g., stellar age, metallicity). The D4000 strength is correlated to the age of stellar populations: as galaxies grow old, they become redder and the strength of D4000 increases. However, it is also correlated with metallicity \citep{Worthey1994}, and when D4000 is used to determine galaxy ages, it is generally combined with metallicity indicators to break the age-metallicity degeneracy.

Given these characteristics, the D4000 has been widely used in galaxy evolution works as a measurement of stellar age, based mostly on D4000$_{\rm n}$ \citep[e.g.][]{Balogh1999, Brinchmann2004, Marcillac2006, Siudek2017}, but also D4000$_{\rm w}$ \citep[e.g.][]{Bruzual1983, Mignoli2005, Kriek2011, Kim2018}, or even alternative definitions \citep{Tresse1999}. In addition to these studies of galaxy properties, the D4000 has also been used in cosmology to independently determine the Hubble expansion rate, $H(z)$, with the method of cosmic chronometers \citep{Moresco2012}. In that work, the measurement of D4000$_{\rm n}$ is used to infer stellar ages of passive red galaxies; the age difference between redshift bins allows us to directly estimate the evolution of the Hubble parameter.

While the D4000 has been traditionally measured with spectroscopic observations, narrow-band photometric surveys may also yield enough spectral resolution for its measurement \citep{Stothert2018, Angthopo2020}. In this work, we will assess the possibility of measuring the D4000 with PAUS narrow-band photometry. In fact, due to its high spectral resolution for an imaging survey, several works have already partially explored the possibilities of its photometry to detect spectral features and/or derive galaxy properties with PAUS. For example, in \citet{Johnston2021} and \citet{Tortorelli2021} \textsc{CIGALE} is applied to PAUS photometry to derive rest-frame colours and luminosities, as well as stellar masses and SFRs, while the photometric redshift code developed in \citet{Alarcon2021} also estimates emission line fluxes. Detection of spectral features has also been evaluated, like the Ly$\alpha$ line for intensity mapping studies in \citet{Renard2021}, or even the possibility of D4000 measurements briefly evaluated with mock catalogues and synthetic datasets in \citet{Stothert2018}, which is a precursor to the work presented here.

In order to evaluate PAUS capabilities to determine the D4000 value, we develop an estimator for directly measuring this spectral feature over its narrow-band photometry, and compare this methodology against the spectroscopic measurements provided by the VIMOS Public Extragalactic Redshift Survey (VIPERS), as well as the reconstruction of the D4000 values obtained from the SED-fitting code CIGALE. For clarity, we will refer throughout this work to the D4000 value measured with PAUS narrow-band photometry as \textit{photometric measurement}, the D4000 measurement over VIPERS spectra as \textit{spectroscopic measurement}, and the value inferred from CIGALE SED fitting as \textit{reconstruction}.

These estimation methods will be compared by evaluating the D4000 distribution and its average trends versus different observables (redshift, stellar mass, SFR). The validity of these D4000 estimations to discriminate between red and blue galaxies when compared to a fiducial classification will also be assessed. It is worth noting that this is the first work where the D4000 is determined by CIGALE using narrow-band photometry and compared to spectroscopy, although other works using CIGALE-derived D4000 with only broad-band photometry already exist. For example, in \citet{Buat2011} it is studied along other spectral features, in \citet{Boquien2012} it is used to examine the dependency of physical parameters on attenuation, and in \citet{Johnston2015} it is employed for galaxy classification. Moreover, the reconstruction of the D4000 (as well as other spectral indices) with SED-fitting techniques has also been evaluated in \citet{Mejia-Narvaez2017} with synthetic J-PAS data.

The paper is structured as follows. In \Cref{sec:Data samples}, we will explain the observational datasets that have been used. The D4000 estimator specifically developed for narrow-band photometry will be discussed in \Cref{sec:Methodology: D4000 estimators}, as well as the reconstruction of D4000 and other galaxy properties with \textsc{CIGALE}. In \Cref{sec:Accuracy and distribution of D4000 estimations with PAUS and CFHTLS data}, the performance of the D4000 estimator will be evaluated against spectroscopic and SED-fitting results by evaluating its SNR, distribution and bias. In \Cref{sec:D4000 and galaxy properties}, the evolution of the D4000 as determined by each estimation method will be examined, as well as its performance for galaxy classification. We will conclude with \Cref{sec:Conclusions}.

\section{Data samples}\label{sec:Data samples}

\subsection{PAUS}\label{sec:PAUS}

\begin{figure}
 	\includegraphics[width=\columnwidth]{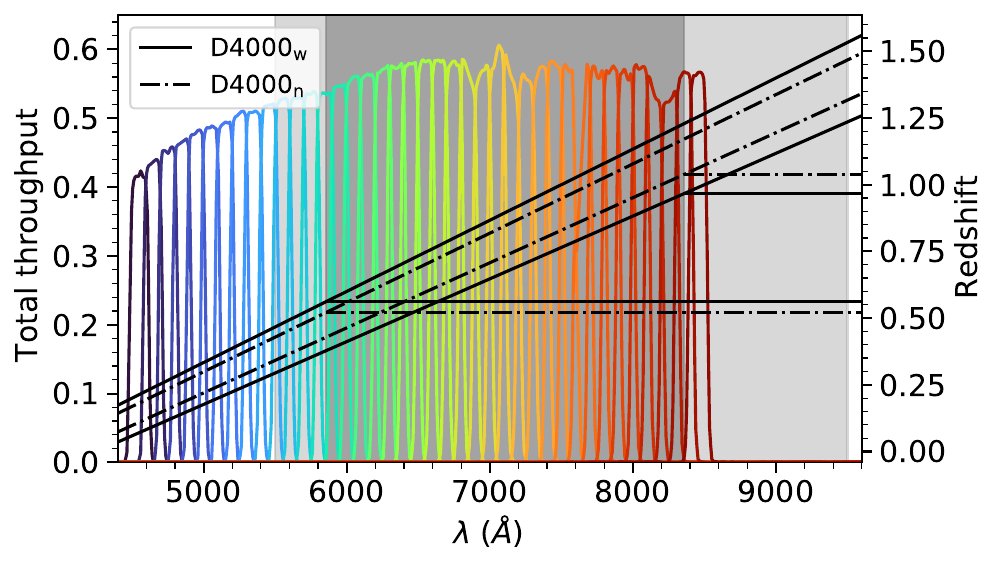}
     \caption{Response functions for the PAUS narrow-band filters (coloured), together with VIPERS total wavelength range (light-grey shaded area), and the wavelength span considered for this study (dark-grey shaded area). The diagonal black lines represent the redshift of the wavelength limits of the D4000 (right y-axis) versus its observed wavelength, while the horizontal black lines highlight the redshift range for the D4000 measurement, given the wavelength coverage of our D4000 study. Solid line for D4000$_{\rm w}$, dash-dotted line for D4000$_{\rm n}$.}
     \label{fig:PAUS_filters}
\end{figure}

PAUS is a narrow-band photometric survey carried out at the William Herschel Telescope. Observations are performed with a special-purpose instrument, the PAU Camera \citep{Padilla2019}, with a configuration of 40 narrow-band filters (NBs) uniformly spaced between 4550 \AA\ and 8450 \AA\ in steps of 100 \AA\ (see \cref{fig:PAUS_filters}). This filter configuration yields an average spectral resolution of $\rm R \sim 65$. It has targeted five different fields: the COSMOS field \citep{Laigle2016} and the wide fields (W1, W2, W3, W4) of the Canada-France-Hawaii Telescope Legacy Survey \citep[CFHTLenS,][]{Cuillandre2012, Hudelot2012}. Photometry in all these fields is performed based on a reference broad-band catalogue (COSMOS and CFHTLS respectively). For each object, an aperture mask is computed using the coordinates of the reference catalogue, and convolving the shape of the object (point source for stars, Sérsic profile for galaxies) with the point spread function (PSF) of the single-epoch image. Fluxes are integrated in these aperture masks up to a certain percentage of the estimated total flux of the object  (a process known as \textit{forced photometry}, see \citealt{Eriksen2019}, \citealt{Cabayol2021} and Serrano et al., in prep.)

The current angular footprint of the survey is $\sim 50$ deg$^2$, and its catalogue is complete up to $i_{\rm AB}\leq23$, with slightly over 3 million galaxies if all fields are considered. For all these galaxies, the computation of photo-z is currently restricted for objects in  the range $0<z<1.2$, hence all objects at $z>1.2$ are placed at lower redshifts in current photo-z runs (work is in progress to extend redshift calculations up to $z<2$). These $z>1.2$ objects accounts for 5\% of the objects approximately, however, it does not affect the results in our paper, as the redshift cuts in our selected sample exclude any objects with spec-z above 1.2 (see \cref{sec:Sample selection}).

Early results show that the photometric redshift is determined with an error of $\sigma_{68}/(1+z) = 0.0037$ to $i_{\rm AB}<22.5$ for a 50\% redshift quality cut \citep{Eriksen2019} in the COSMOS field. These benchmark results have been improved with enhancements over the original photo-z code \citep{Alarcon2021}, or machine learning approaches \citep{Eriksen2020, Soo2021}. In all these works (as well as in this paper), the photo-z is computed using both PAUS NBs and the broad-band photometry of the reference catalogues (CFHTLS for the W1 field).

\subsection{VIPERS}\label{sec:VIPERS}

The VIMOS public extragalactic redshift survey \citep[VIPERS,][]{Guzzo2013, Scodeggio2018} is a spectroscopic survey carried out at the Very Large Telescope in Cerro Paranal, Chile, with the VIMOS instrument \citep{LeFevre2003}. The grism used for this survey (low-resolution red) yields a spectral resolution $R \simeq 220$ in a wavelength range of 5500-9500 \AA\ (light-grey shaded area in  \cref{fig:PAUS_filters}) For this work, we utilise its second public data release, PDR-2 \citep{Scodeggio2018}.

The target selection of PDR-2 comprises objects in the redshift interval $0.5<z<1.2$, with a magnitude limit of $i_{\rm AB}\leq 22.5$, which makes it one of the deepest spectroscopic catalogues publicly available.  It is worth noting that deeper spectroscopic surveys do exist in fields overlapping with PAUS, such as VANDELS \citep{Pentericci2018}, VUDS \citep{LeFevre2015}, VVDS \citep{LeFevre2013} or zCOSMOS-deep \citep{Lilly2006}, just to mention the surveys employing the same spectrograph as VIPERS. However, VIPERS has the largest angular coverage of them, as well as the closest magnitude limit to PAUS ($i_{\rm AB}<22.5$ vs $i_{\rm AB}<23$). Moreover, at its redshift range ($0.5<z<1.2$), VIPERS also offers an unique combination of sampled volume and galaxy density \citep[almost 90,000 galaxies, see][]{Scodeggio2018}. Hence, it can be considered the closest catalogue to a perfect PAUS spectroscopic counterpart. The complete catalogue covers a total of $\sim 23.5$ deg$^2$, split between the CFHTLS W1 and W4 fields, which results in a catalogue of 86,775 galaxies in total.

\subsection{Sample selection}\label{sec:Sample selection}

Thanks to its unprecedented volume and the wealth of auxiliary data, VIPERS is an ideal spectroscopic dataset to validate PAUS capabilities. We have cross-matched the W1 VIPERS catalogue with PAUS observations, using a radius of 1 arcsec to match objects between catalogues. We have not considered the W4 field, as the current coverage of PAUS in W4 is significantly smaller than W1 (approximately 2 deg$^2$ vs 14 deg$^2$), and very uneven between narrow bands. For these reasons, no photo-z calculation in W4 has been carried out yet at the time of writing.

\begin{figure*}
 	\includegraphics[width=\textwidth]{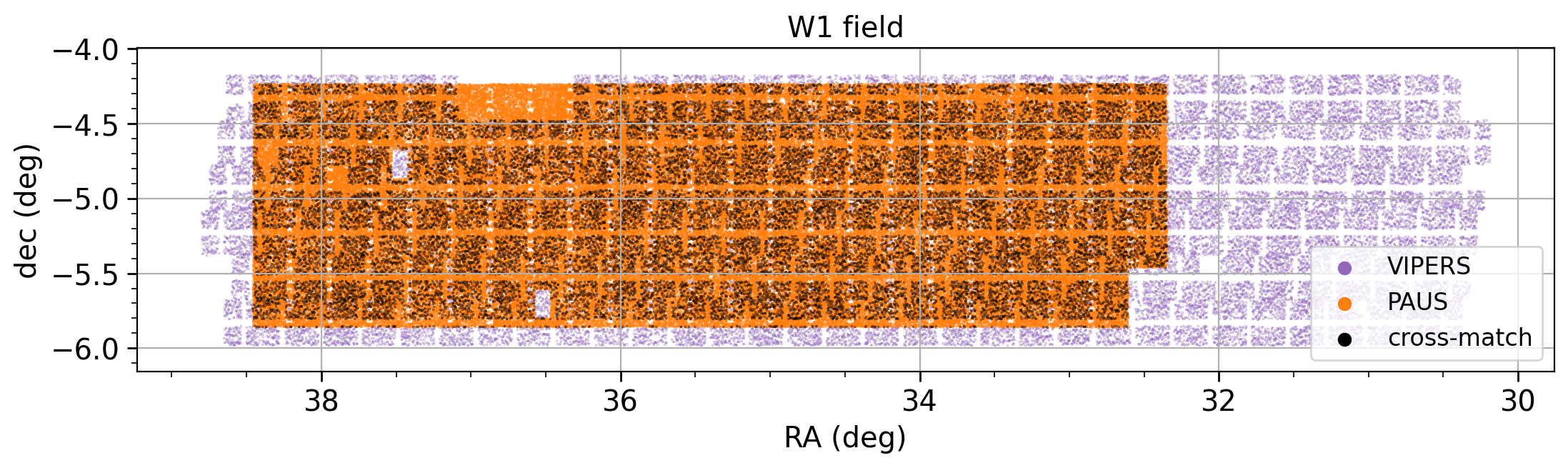}
     \caption{Survey footprint in the W1 field for PAUS (orange), VIPERS (purple) and the cross-match between both (black). Each point represents an object in the catalogue.}
     \label{fig:W1_angular_footprint}
\end{figure*}

In \cref{fig:W1_angular_footprint}, the footprints of both catalogues in the W1 field are displayed; the cross-match between both yields a sample of 33,363 objects covering the whole PAUS redshift range ($0<z<1.2$). Moreover, we applied a cut based on the VIPERS redshift quality flags: only objects with $2\leq \texttt{zflg} <10$ or $22\leq \texttt{zflg} <30$ were included in the sample \citep{Scodeggio2018}. This cut keeps all the primary (i.e., in the original target catalogue) and secondary (i.e., observed in the slit but not in the target catalogue) objects with a redshift measurement good enough for science (>90\% confidence level), while removing AGNs from the sample. This redshift quality cut leaves the selected sample with 28,788 galaxies.

Given that VIPERS spectra are observed in a redder wavelength span than PAUS, the wavelength range selected for D4000 measurements goes from 5850 \AA\ to 8350 \AA\ (\cref{fig:PAUS_filters}, dark-grey area). With this choice of wavelength coverage, the low SNR regions at the limits of the spectrograph range are avoided, and an extra PAUS NB is left outside of range in the red end to avoid undesired boundary effects in the D4000 estimation. This restriction in wavelength coverage also implies a constraint in the redshift interval where the D4000 can be measured, as shown in \cref{fig:PAUS_filters} by the horizontal black lines. For D4000$_{\rm w}$ and D4000$_{\rm n}$, with wavelength ranges defined in \cref{tab:D4000_wavelengths}, this redshift range is $0.562<z<0.967$ and $0.521<z<1.039$ respectively. Nevertheless, we have applied just the most stringent cut of $0.562<z<0.967$, in order to use the same sample to evaluate both D4000$_{\rm w}$ and D4000$_{\rm n}$. After applying this redshift cut to both PAUS photo-z and VIPERS spec-z, the selected sample has 17,375 objects, with mean redshifts $\langle z_{\rm PAUS} \rangle=0.731$ and z $\langle z_{\rm VIPERS} \rangle = 0.743$ (see \cref{fig:z_comparison}). This slight discrepancy in mean redshift may be due to the redshift focusing artefacts in the PAUS photo-z distribution, which appear as horizontal bands in \cref{fig:z_comparison}. This effect is stronger for VIPERS objects at $z>0.7$, where some galaxies are wrongly assigned a PAUS photo-z in the range $0.7<z<0.85$, thus lowering the mean redshift of the whole distribution. Nevertheless, despite these artefacts the overall PAUS photo-z performance is significantly better than the CFHTLS broad-band photo-z, as discussed in \cref{sec:Redshift comparison}.

In addition to the redshift cut, once the D4000 has been computed all objects with an invalid D4000 measurement have been removed. This includes objects with missing NBs or masked spectral regions needed for the photometric D4000 calculation, as well as objects whose CIGALE fits were deemed not good enough. This leaves a final sample of 17,241 galaxies that is used throughout this work.

\section{Methodology: D4000 estimators}\label{sec:Methodology: D4000 estimators}

In this section, we will describe in detail the two different methods used to measure the D4000 from PAUS data: the photometric measurement by determining the D4000 rest-frame bands using PAUS NBs (\cref{sec:Direct estimator}), and the reconstruction with the CIGALE SED-fitting that, among other physical parameters, yields the D4000 of the galaxies in the selected sample (\cref{sec:CIGALE estimation}). Moreover, in \cref{sec:Direct estimator} we will also briefly review the general D4000 definition, as well as its calculation with spectroscopic data.

\subsection{Direct estimator}\label{sec:Direct estimator}

\subsubsection{Estimator definition}\label{sec:Estimator definition}

{The measurement of the D4000} is defined as the ratio between two average flux frequency densities ($F_\nu$), integrated in rest frame (also known as \textit{rest-frame bands}), 

\begin{equation}\label{eq:D4000_definition}
{\rm D4000} = \frac{\langle F_\nu \rangle_{\rm red}}{\langle F_\nu \rangle_{\rm blue}}.
\end{equation}

If these average flux density bands are to be directly measured on spectra, its integration is straightforward,

\begin{equation}\label{eq:D4000_band_integration}
\langle F_\nu \rangle = \frac{1}{(\lambda_{\rm max} - \lambda_{\rm min})(1+z)} \int_{\lambda_{\rm min}\cdot(1+z)}^{\lambda_{\rm max}\cdot(1+z)} d\lambda\, F_\nu(\lambda),
\end{equation}
where $\lambda_{\rm min}$ and $\lambda_{\rm max}$ are the wavelength boundaries for a given band, and $z$ is the redshift of the object.

In \cref{eq:D4000_definition}, $\langle F_\nu \rangle_{\rm red}$ is the integrated flux of the rest-frame band defined at wavelengths larger than 4000\AA\ (i.e., the region unaffected by the combination of absorption lines that causes the break itself), and $\langle F_\nu \rangle_{\rm blue}$ is the integrated flux at the rest-frame band at shorter wavelengths (where the continuum is affected by the set of absorption lines of the 4000 \AA\ break). The wavelength boundaries of these rest-frame bands are specified in \cref{tab:D4000_wavelengths}, as specified in \citet{Bruzual1983} for D4000$_{\rm w}$ and \citet{Balogh1999} for D4000$_{\rm n}$. We will refer to these bands, as well as to their integrated fluxes, as \textit{red band} and \textit{blue band} respectively (or simply \textit{D4000 bands}). The stronger the spectral break (i.e., the more absorption), the fainter the blue band compared to the red one, and the higher the D4000 value (since in \cref{eq:D4000_definition} the flux of the red band is in the numerator, while the blue band flux is in the denominator).

In this work, the spectroscopic measurements of the D4000 from VIPERS spectra have been computed following \cref{eq:D4000_band_integration} with the EZ redshift code \citep{Garilli2010}. If the D4000 has to be determined with narrow-band photometry, however, \cref{eq:D4000_band_integration} cannot be directly applied. The spectral resolution of the NBs is of the same order of magnitude as the width of the D4000 bands, and thus common numerical integration methods (e.g., Trapezoidal rule, Simpson's rule) may result in large inaccuracies. Consequently, we have developed a specific estimator to compute the D4000 bands with narrow-band photometry. While its performance will only be evaluated for PAUS in this work, by definition it can be applied to any combination of narrow-band filters, as long as their FWHM is smaller than the D4000 bands when shifted to the observed frame, and the wavelength coverage in the redshift range of interest is reasonably complete.

For PAUS NBs, $\langle F_\nu \rangle$ is computed as

\begin{equation}\label{eq:PAUS_estimator}
\langle F_\nu \rangle  = \frac{\sum_{\rm i}\Delta \lambda_{\rm i} r_{\rm i}^2 \sigma_{\rm i}^{-2} F_{\rm\nu\,i}}{\sum_{\rm i} \Delta \lambda_{\rm i} r_{\rm i}^2 \sigma_{\rm i}^{-2}},
\end{equation}
where $F_{\rm \nu\, i}$ and $\sigma_{\rm i}$ are the flux and its error for the band $i$, respectively, and the sum is over all the NBs considered for the photometric D4000 measurement (\cref{fig:PAUS_filters}).

The $r_{\rm i}$ factor is the following ratio

\begin{equation}\label{eq:r_definition}
r_{\rm i} = \frac{\int_{\lambda_{\rm min}\cdot(1+z)}^{\lambda_{\rm max}\cdot(1+z)}d\lambda\, R_{\rm i}(\lambda)}{\int_0^\infty d\lambda\, R_{\rm i}(\lambda)},
\end{equation}
where $R_{\rm i}(\lambda)$ is the response function for the filter $i$. In the numerator, $R_{\rm i}(\lambda)$ it is integrated only over the limits of the D4000 band of interest (as defined in \cref{tab:D4000_wavelengths}), redshifted to the observed frame (hence the redshift of the object, $z$). In the denominator, however, the total area of $R_{\rm i}(\lambda)$ is computed. Thus, this $r_{\rm i}$ factor can be interpreted as the fraction of the NB that is inside the wavelength range of the D4000 band.

Regarding the $\Delta \lambda_{\rm i}$ factor in \cref{eq:PAUS_estimator}, it must be a term with wavelength units to make the broader NBs weight more in the calculation of the D4000 bands. The definition we have chosen is simply

\begin{equation}\label{eq:d_lambda_definition}
\Delta \lambda_{\rm i} = \int_0^\infty d\lambda\, R_{\rm i}(\lambda),
\end{equation}

If the response functions $R_{\rm i}(\lambda)$ were ideal top-hat functions with a maximum value of 1, intuitively $\Delta \lambda_{\rm i}$ would be the width of the top-hat, and the product $\Delta \lambda_{\rm i}F_{\rm \nu\, i}$ would simply correspond to the area of the flux integrated by the NB $i$. Consequently, \cref{eq:PAUS_estimator} in this ideal case could be interpreted as a Riemann sum weighted by $r_{\rm i}^2 \sigma_{\rm i}^{-2}$. Hence, \cref{eq:PAUS_estimator} is the generalisation of a weighted Riemann sum to any set of photometric bands with response functions $R_{\rm i}(\lambda)$.

The weighting $r_{\rm i}^2 \sigma_{\rm i}^{-2}$ can be interpreted as the contribution of two different error sources: the error of the NB measurement itself ($\sigma_{\rm i}$), and the part of the NB corresponding to the flux integrated outside of the D4000 band of interest. All flux information coming from outside the D4000 band wavelength range can be regarded as noise; therefore a NB which has half of its area outside of the D4000 band should have half the SNR than a NB which is fully inside. This can be expressed in terms of a net error, $\sigma_{\rm net\, i} = r_{\rm i}^{-1} \sigma_{\rm i}$. Therefore, the weighting is just a standard inverse variance weighting ($\sigma_{\rm net\, i}^{-2}$), but using a $\sigma_{\rm net\, i}$ error that accounts for both the intrinsic error of the observation of each band flux, and the undesired flux information outside of the D4000 rest-frame band. Regarding the error of \cref{eq:PAUS_estimator}, it is simply determined by standard error propagation of the errors of the NB fluxes, $F_{\rm \nu \, i}$

\begin{figure}
 	\includegraphics[width=\columnwidth]{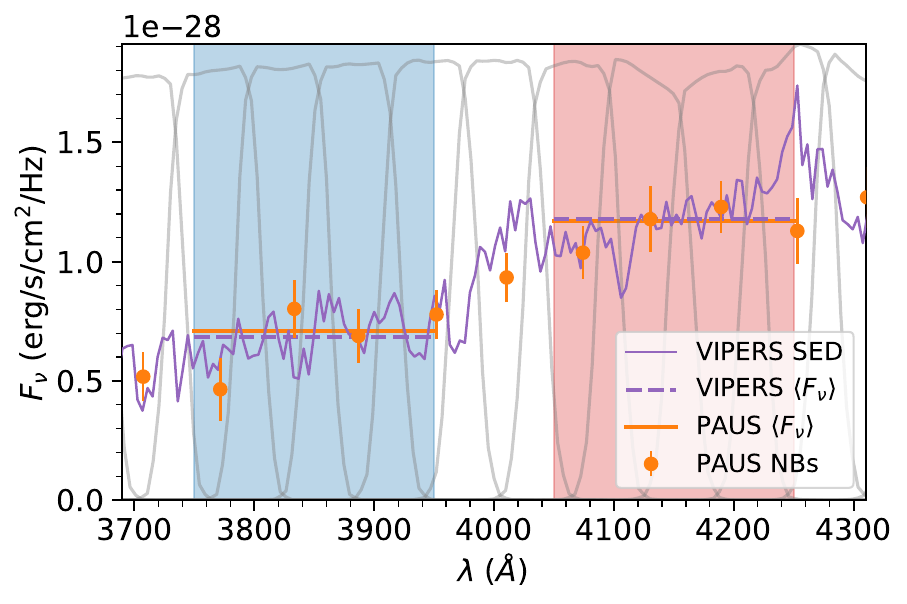}
     \caption{Example of photometric and spectroscopic measurement of D4000$_{\rm w}$ bands (red and blue areas). PAUS NB fluxes are represented by orange points with error bars, while the VIPERS spectrum is shown with a solid purple line; the horizontal orange solid line and dashed purple line represent their respective D4000$_{\rm w}$ bands. PAUS response functions in rest-frame are plotted in very light grey for reference (normalised to fit the plot). The VIPERS spectrum has been shifted to the rest frame using its spectroscopic redshift; while the PAUS photo-z has been used for PAUS data.}
     \label{fig:D4000_measurement_example}
\end{figure}

In \cref{fig:D4000_measurement_example} an example of the D4000$_{\rm w}$ photometric (with PAUS NBs) and spectroscopic (with the VIPERS spectrum) measurement for a single object is displayed. The selected object is a galaxy with D4000$_{\rm w} \sim 1.7$, a value high enough to have a spectral break easily distinguishable by eye, and typical for red galaxies. Hence, it can be considered a fairly representative example of a red galaxy in our sample. The apparent magnitude of the object is $i_{\rm AB}=20.71$. The plot is shown in rest frame, and the values of the D4000$_{\rm w}$ band fluxes are determined with \cref{eq:D4000_band_integration} for the spectroscopic measurement and \cref{eq:PAUS_estimator} for the photometric measurement, respectively. Besides, the PAUS NB response functions are plotted with a faint grey colour for reference. The relevant data of regarding the D4000 measurement of the object shown in \cref{fig:D4000_measurement_example} are displayed in \cref{tab:example_object_data}. It is worth noting that if the PAUS D4000 is computed using the VIPERS spec-z, we find a photometric D4000$_{\rm w}=1.50\pm 0.14$, a result consistent with the PAUS value displayed in \cref{tab:example_object_data}, but not so much with its VIPERS counterpart. The reason for this discrepancy may lie in how the noise in PAUS NBs influences the determination of each photo-z, yielding D4000 values that are more self-consistent than these computed with an external redshift measurement. However, as it is shown in \cref{sec:SNR}, the actual D4000 error budget for photometric measurements is perfectly accounted for by the errors in PAUS photometry alone.

\begin{table}
\centering
	\caption{Wavelength ranges for the blue and red rest-frame bands of \cref{eq:D4000_definition}, for both the D4000$_{\rm w}$ and D4000$_{\rm n}$ definitions.}
 	\label{tab:D4000_wavelengths}
 	\begin{tabular}{rcc}
 		\hline
 		 & Blue band (\AA) & Red band (\AA)\\
 		\hline
 		D4000$_{\rm w}$  & 3750-3950 & 4050-4250\\
        D4000$_{\rm n}$  & 3850-3950 & 4000-4100\\
		\hline
 	\end{tabular}
 \end{table}

\begin{table}
\centering
	\caption{Redshifts, D4000$_{\rm w}$ measurements and catalogue IDs for the representative object displayed in \cref{fig:D4000_measurement_example}. The error for VIPERS spec-z is the average error specified in \citet{Scodeggio2018}.}
 	\label{tab:example_object_data}
 	\begin{tabular}{rccc}
 		\hline
 		 & ID  & z & D4000$_{\rm w}$ \\
 		\hline
        VIPERS & 117158758 & 0.6995 $\pm$ 0.0013 & 1.73 $\pm$ 0.03 \\
 		PAUS photo-z & \multirow{2}{*}{10691571} & 0.66 $\pm$ 0.025 & 1.65 $\pm$ 0.18\\
  		PAUS spec-z &  & 0.6995 $\pm$ 0.0013 & 1.50 $\pm$ 0.14\\
		\hline
 	\end{tabular}
 \end{table}

\subsubsection{Synthetic PAUS tests}\label{sec:Synthetic PAUS tests}

Before applying it to real PAUS data, the photometric estimator for the D4000 bands defined in \cref{eq:PAUS_estimator} has been tested against synthetic PAUS NBs, in order to evaluate the potential biases stemming from the estimator alone, without the experimental errors in the PAUS photometry or photo-z, as well as other possible unaccounted systematics.

The synthetic PAUS dataset (sPAUS) has been generated by selecting the whole VIPERS sample in W1, and integrating the spectra for all PAUS NBs considered in this work (\cref{fig:PAUS_filters}). These synthetic NBs have been used to determine the D4000 with \cref{eq:PAUS_estimator}, using the respective spectroscopic VIPERS redshifts. Therefore, any deviations from the spectroscopic measurements must be due to our photometric estimation method of the D4000 (which is limited by the spectral resolution of the narrow-band photometry).

\begin{figure*}
 	\includegraphics[width=\columnwidth]{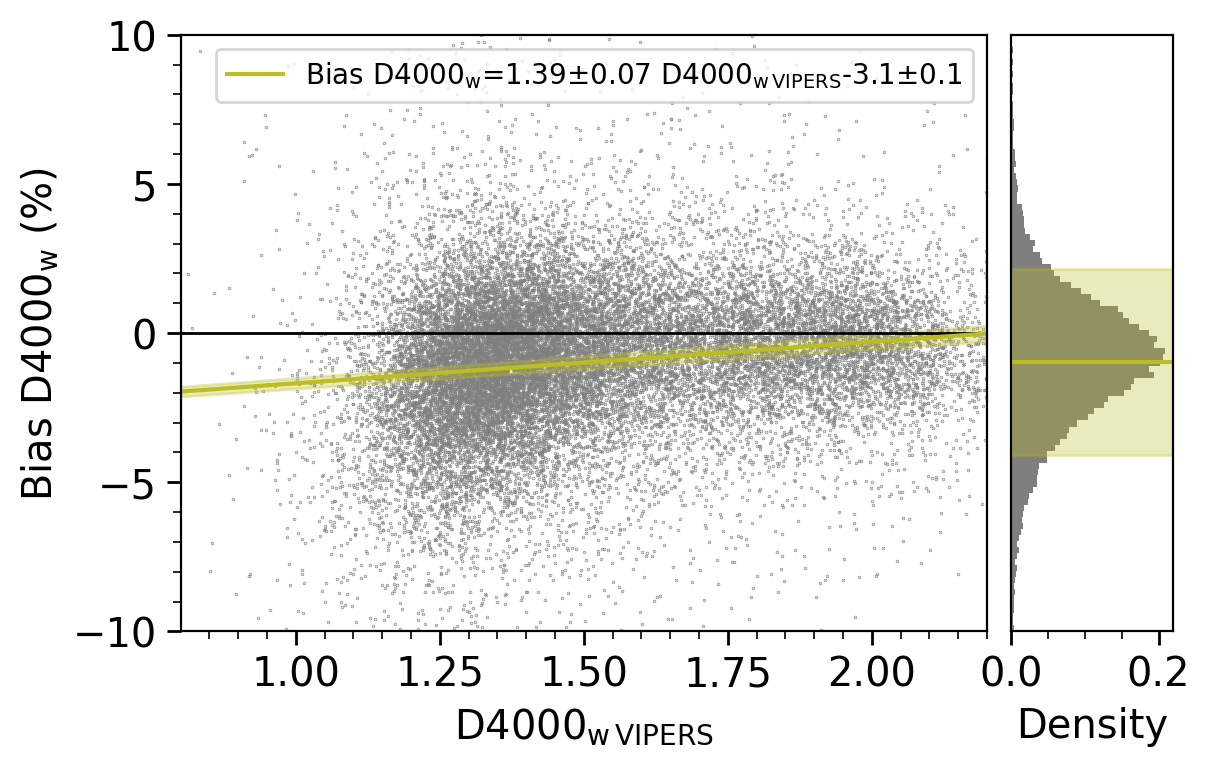}
 	\includegraphics[width=\columnwidth]{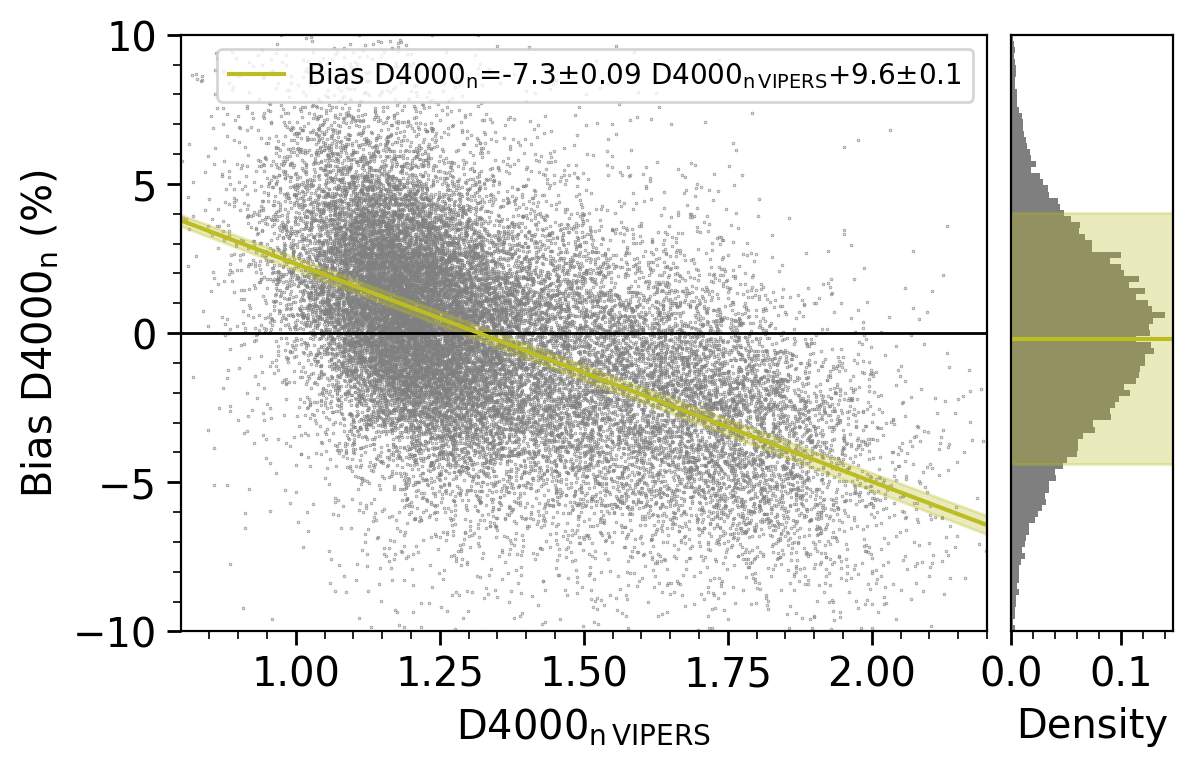}
     \caption{The bias of the photometric D4000 measured in the synthetic PAUS NBs, vs VIPERS D4000 values. A linear fit with its error and the mean bias and its $\sigma$ in the histograms are marked with yellow lines and areas, respectively. \textit{Left:} D4000$_{\rm w}$. \textit{Right:} D4000$_{\rm n}$.}
     \label{fig:bias_scatter_synth}
\end{figure*}

A comparison between the VIPERS D4000 and the sPAUS D4000 is shown in \cref{fig:bias_scatter_synth}, both for D4000$_{\rm w}$ and D4000$_{\rm n}$. In both panels, the scatter plot on the left displays the VIPERS D4000 versus the bias of the sPAUS D4000, defined as 

\begin{equation}\label{eq:bias_definition}
\rm{Bias\, (\%)} = 100 \cdot\left( \frac{D4000_{\rm sPAUS}}{D4000_{\rm VIPERS}} - 1 \right),
\end{equation}

together with a linear fit and its error (yellow line). The histogram on the right is a histogram of the bias, with its mean value and $\sigma$ shown in yellow. From these plots, it can be inferred that the sPAUS D4000$_{\rm w}$ has a slight mean negative bias of the order of $\sim$-1\% that seems fairly constant across all the VIPERS D4000 range (the linear fit shows just a small correlation). On the other hand, the sPAUS D4000$_{\rm n}$ shows a smaller net bias but also more dispersion, and a clear correlation between bias and spectroscopic D4000 values: the SPAUS D4000$_{\rm n}$ is overestimated for bluer galaxies (D4000$_{\rm n}<1.4$), while it is underestimated for higher values (i.e.,  for redder galaxies). Moreover, in both cases the total bias distribution in the histograms is fairly close to Gaussian. Some numerical results from this comparison are summarised in \cref{tab:bias_results}.

Overall, these results show a significant improvement over the preliminary analysis carried out in \citet{Stothert2018}, where a simple linear interpolation was used to estimate the photometric D4000 over a similar synthetic catalogue (computed from SDSS spectra instead of VIPERS). In Figure 10 of \citet{Stothert2018}, a comparison akin to \cref{fig:bias_scatter_synth} is displayed, but showing only mean biases (without \%) in bins of the VIPERS D4000. While the same trends appear (a clear linear correlation in the D4000$_{\rm n}$), the absolute bias values are significantly lower in this work. This is especially noticeable on the low and high D4000 ends, e.g., at D4000$_{\rm n}\sim 2 $ \citet{Stothert2018} finds a mean bias of $\sim$-7.5\% (dashed line), while in this work the linear fit shows a bias of $\sim$-5\%. A similar improvement appears in the D4000$_{\rm w}$, where they find biases of $\sim$-7.5\% and $\sim$-2.5\% for the low and high D4000 ends, while \cref{fig:bias_scatter_synth} clearly shows smaller mean biases in these regions. Besides, their mean bias of $\sim$-2\% for the photometric D4000$_{\rm w}$ found in \citet{Stothert2018} is reduced by a factor of two.

The reason for this improvement is the integration of the D4000 bands over NBs using an estimator that properly weights the flux fraction in each NB that is relevant to the rest-frame band (\cref{eq:PAUS_estimator}), as well as the observational error. With linear interpolation, NBs mostly outside the rest-frame band can still affect the slope of the interpolated SED (if they are adjacent to NBs "fully" inside the rest-frame band), and thus affect significantly the value of the integrated flux (\cref{fig:D4000_measurement_example} may help visualising this effect).

\begin{table}
\centering
	\caption{Mean and $\sigma$ of the bias for the photometric D4000$_{\rm w}$ and D4000$_{\rm n}$ in the synthetic PAUS catalogue, together with the percentage of objects below absolute bias thresholds of 1\%, 2.5\% and 5\%.}
 	\label{tab:bias_results}
 	\begin{tabular}{rccccc}
 		\hline
 		 & & & \multicolumn{3}{c}{\% with absolute bias}\\
 		 & Mean bias (\%)  & $\sigma$ bias (\%) & $<$1\% & $<$2.5\% & $<$5\%\\
 		\hline
 		D4000$_{\rm w}$ & -1.03 & 2.59 & 41.06 & 77.21 & 95.11 \\
 		D4000$_{\rm n}$ & -0.23 & 3.19 & 32.07 & 67.44 & 92.62 \\
		\hline
    \end{tabular}
\end{table}

In addition to the dependency of the sPAUS D4000 bias with the VIPERS D4000, we investigate its correlation with redshift. Given that the D4000 bands are defined in rest frame, the $r_{\rm i}$ factor in \cref{eq:r_definition} and thus the NBs used in the photometric D4000 calculation will be a function of $z$. Therefore, one could expect the redshift regions where the NBs better match the D4000 bands to be less biased and vice-versa (i.e., when the D4000 bands are covered solely by NBs fully inside the band, the measurement should be more accurate). As a consequence, the bias as a function of redshift should exhibit an oscillatory pattern as redshift increases and the D4000 bands shift to redder wavelengths in observed frame (since there will be an alternation of "good" and "bad" NB matches to the D4000 bands).

It is worth noting that this redshift dependency is just an effect of the relative position of the NBs with respect to the redshifted D4000 bands; any possible effect derived from galaxy evolution or observational systematics dependent on redshift is not considered. Since we are working with synthetic NBs integrated from VIPERS spectra, we can shift the wavelength of the spectra with the following change of variable before NB integration,

\begin{equation}\label{eq:apparent_redshift}
    \lambda' = \frac{1+z_{\rm apparent}}{1+z}\lambda,
\end{equation}

where $z$ is the actual redshift of the object, and $z_{\rm apparent}$ is the redshift we want the NBs to "observe" the object with. This transformation is akin to shifting the object to the rest frame and then moving it to a different observed frame defined by $z_{\rm apparent}$. Hence, we can evaluate the redshift bias of any object not only at its real redshift $z$, but at any $z_{\rm apparent}$, as long as the photometric D4000 can still be measured in the new observed frame.

\begin{figure}
 	 \includegraphics[width=\columnwidth]{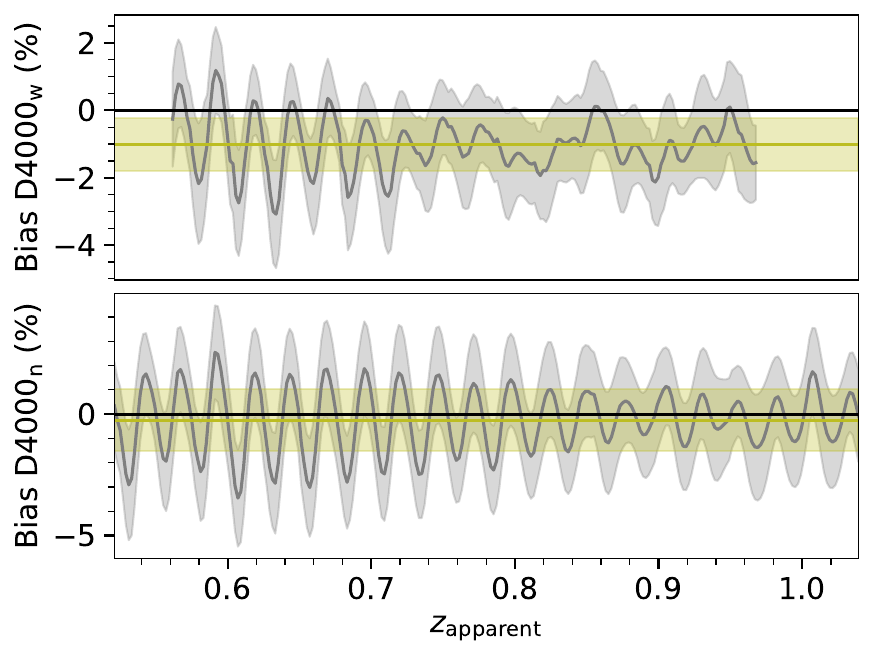}
     \caption{Bias versus $z_{\rm apparent}$ for the sPAUS D4000$_{\rm w}$ (upper panel) and sPAUS D4000$_{\rm n}$ (lower panel). Grey line and shaded area represent the mean bias and its $\sigma$ for each $z_{\rm apparent}$, while the yellow line and its shaded area are the mean bias and $\sigma$ for the whole $z_{\rm apparent}$ range.}
     \label{fig:redshift_bias}
\end{figure}

Consequently, we have evaluated the redshift bias for all objects in the VIPERS W1 sample by computing their NBs in a fine $z_{\rm apparent}$ grid. The results are shown in \cref{fig:redshift_bias}. The bias appears as a continuous function because of the small grid size. At each point, the displayed bias and its error are respectively the mean bias and its $\sigma$ of all the objects of the VIPERS W1 sample where the photometric D4000 could be measured after applying \cref{eq:apparent_redshift}.

For the sPAUS D4000$_{\rm w}$, this redshift bias oscillates $<$2\% around the mean bias, while in the sPAUS D4000$_{\rm n}$ the oscillations reach well above 2\%. In both cases, the oscillations become significantly smaller as redshift increases, because the larger the redshift, the wider the D4000 bands in observed frame, and the smaller the error induced by NBs partially outside the D4000 band ($0<r_{\rm i}<1$). Moreover, the oscillations seem fairly stable around the mean bias (yellow lines); hence we can consider the redshift oscillations independent of the D4000 bias (see \cref{fig:bias_scatter_synth}).

After these tests, we can state that our estimator for NB photometry has a mean bias of $-1.03\pm2.59$\% and $-0.23\pm3.19$\% for photometric D4000$_{\rm w}$ and D4000$_{\rm n}$ respectively (see \cref{tab:bias_results}), plus an oscillatory redshift bias due to the relative position of the NBs to the D4000 bands in observed frame. This oscillatory bias  adds an extra $\sim$2\% to the D4000$_{\rm w}$ in the worst cases (yielding a much smaller effect in most of the redshift range), while this increase is also larger for the D4000$_{\rm n}$. Besides, for the photometric D4000$_{\rm n}$ there is a clear decreasing linear correlation between bias and D4000 value, while for the photometric D4000$_{\rm w}$ this correlation is much weaker. In any case, a significant fraction of the sample shows sub-percent bias (41\% and 31\% for D4000$_{\rm w}$ and D4000$_{\rm n}$ respectively), and the large majority has an absolute bias below 5\% (95\% and 93\% respectively, see \cref{tab:bias_results}). Given these results, no bias corrections have been applied to the photometric D4000 measurements, since the errors of the bias are larger than the biases themselves.

Overall, the estimation of the photometric D4000$_{\rm w}$ through this method seems more reliable, but given the preponderance in the literature of the D4000$_{\rm n}$, we will consider both for scientific use. Considering that these results are for a synthetic dataset generated from spectroscopic data, without the larger photometric and redshift errors from narrow-band photometry, it is safe to assume that the biases of the estimator are negligible, or at least completely subdominant, when compared to the observational errors that we can expect.

\subsubsection{Redshift bias versus filter FWHM}\label{sec:Redshift bias versus filter FWHM}

In addition to analysing the photometric D4000 bias for PAUS NBs, we have further explored the oscillatory redshift bias by repeating the analysis displayed in \cref{fig:redshift_bias} for different filter configurations. Instead of using PAUS NBs (\cref{fig:PAUS_filters}), we have defined a set of top-hat filters of different FWHM (between 210 \AA\, and 50 \AA\, approximately) fully covering the same wavelength range as PAUS without overlap between them. For each FWHM, the synthetic NBs have been determined for all objects, as well as the photometric D4000 and its mean redshift bias, by applying the redshift translation defined in \cref{eq:apparent_redshift}.

By definition, for these ideal filter top-hat sets the filter separation and FWHM are the same; therefore, when we refer to the FWHM of the top-hat filters we also refer to the separation between them. For PAUS however, filter separation is $\sim$100 \AA\, while their FWHM is $\sim$130 \AA. In a realistic setting, the response function of a filter will never be a top-hat function, and some filter overlap will be inevitable if a reasonably full coverage is to be expected. Regardless, this simple study with top-hat filters may be useful to evaluate if different filter separations may yield less biased D4000 measurements.

\begin{figure*}
 	\includegraphics[width=\textwidth]{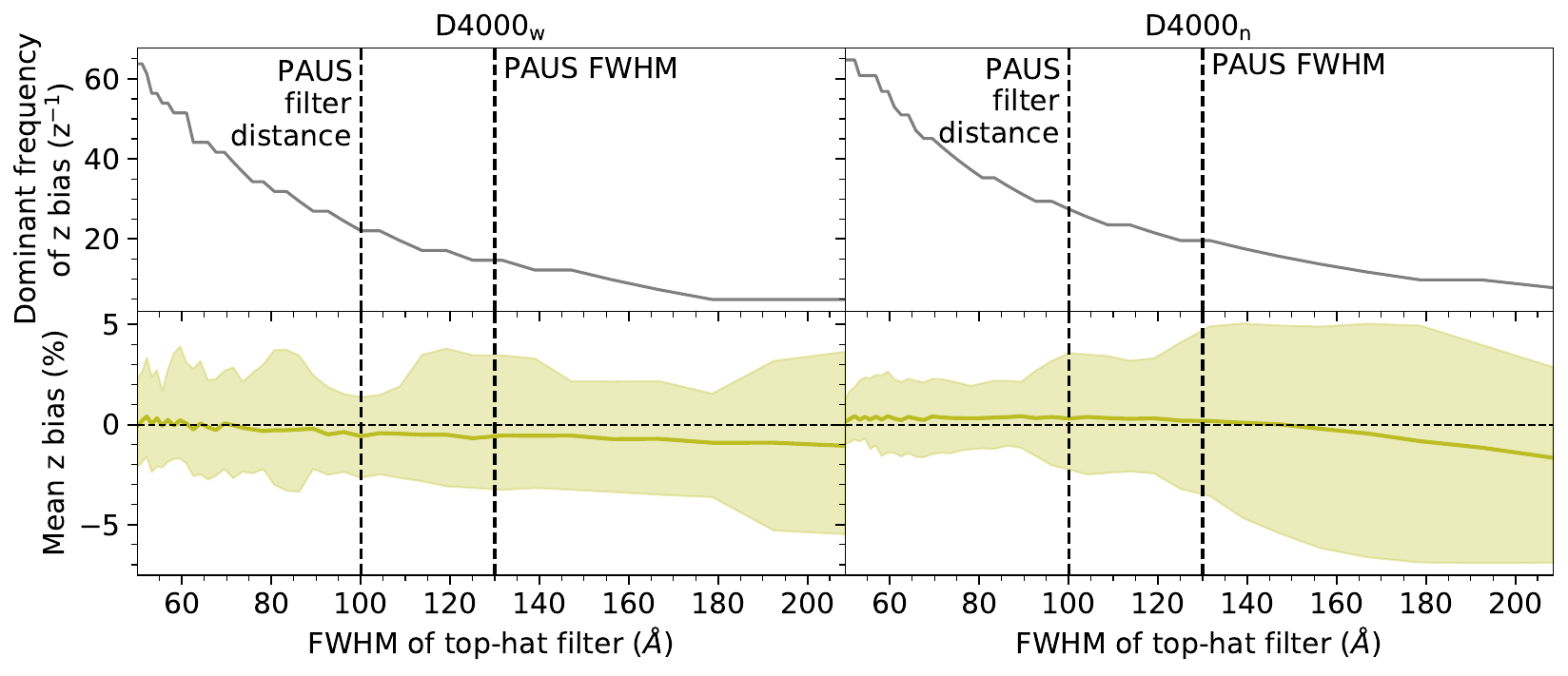}
     \caption{Study of redshift bias versus FWHM of a set of identical top-hat filters fully covering without overlap the PAUS wavelength range, evaluated in the same redshift ranges as \cref{fig:redshift_bias}. Top row displays the dominant frequency of the redshift bias, bottom row shows the mean bias across the redshift range (yellow line), together with its maximum and minimum value (shaded area). Left column for D4000$_{\rm w}$, right column for D4000$_{\rm n}$. Vertical dashed lines represent the approximate filter separation and filter FWHM for PAUS. The y-axis scale is shared between the D4000$_{\rm w}$ and D4000$_{\rm n}$ plots.}
     \label{fig:redshift_bias_vs_fwhm}
\end{figure*}

\cref{fig:redshift_bias_vs_fwhm} shows the dominant frequency of the redshift bias versus FWHM, determined via fast Fourier transform. Moreover, the mean, maximum and minimum value of the redshift bias vs FWHM is also evaluated (for the same redshift range as \cref{fig:redshift_bias}). The mean is the same as the yellow line in \cref{fig:redshift_bias}, while the maximum and minimum are the highest peak and lowest valley of the oscillatory grey line in \cref{fig:redshift_bias}, respectively. Both D4000$_{\rm w}$ and D4000$_{\rm n}$ show a very similar frequency decrease with FWHM, which seems to asymptotically approach zero as the FWHM of the filters becomes significantly larger than the width of the D4000 bands (\cref{tab:D4000_wavelengths}). The mean redshift bias is kept at sub-percent levels for the whole FWHM range for D4000$_{\rm w}$, while for D4000$_{\rm n}$ this is only true at $\rm FWHM<180$ \AA\, (after a drastic increase in the absolute bias value at $\rm FWHM>150$ \AA). Regardless, it seems that the average D4000$_{\rm w}$ bias can only be improved by reducing the FWHM up to 80 \AA, while for D4000$_{\rm n}$ the current PAUS FWHM already yields an optimal result.

On the other hand, the maximum and minimum redshift biases fluctuate more erratically with the FWHM, as the redshift bias signal is composed of different frequencies (this can be seen in \cref{fig:redshift_bias}, although for the PAUS case the non-dominant frequency components are not very noticeable). For the case of D4000$_{\rm w}$, there seems to be a sweet spot around $\rm FWHM \sim 100$ \AA\, where the redshift bias is contained in the smallest range possible (around $\pm$2\%). For D4000$_{\rm n}$, $\rm FWHM < 90$ \AA\ provides the smallest redshift bias oscillation, and using even narrower filters does not seem to decrease the bias range. Given that the PAUS filter separation is already 100 \AA, the decrease we may expect in D4000 photometric measurement with even narrower bands will be fairly limited. In fact, as we will see in \cref{sec:SNR}, the noise in band fluxes will be the limiting factor for the photometric D4000 measurement, so the usage of narrower bands may be indifferent or even counterproductive if exposure times are not increased accordingly.

\subsection{CIGALE estimation}\label{sec:CIGALE estimation}

In addition to the direct measurement over PAUS NBs, we have used the SED fitting algorithm Code Investigating GALaxy Emission \citep[CIGALE;][]{Noll2009, Boquien2019} to perform a D4000 reconstruction from photometric observations. CIGALE is a physically-motivated state-of-the-art Python code for SED fitting based on the principles of the energetic balance between dust-absorbed stellar emission and its re-emission in the infrared (IR). The capabilities of this SED fitting tool have already been verified on PAUS observations \citep{Johnston2021, Tortorelli2021}. For this work, we have used a delayed star formation history, \citet{Bruzual2003} single stellar population models with the initial mass function (IMF) given by \citet{Chabrier2003}, a \citet{Charlot2000} attenuation law and the dust emission models of \citet{Dale2014} to built a grid of models. A detailed description of each module can be found in \citet{Malek2018} and \citet{Boquien2019}.

The adopted parameters employed for the SED fitting are listed in \cref{tab:cigale}.
Thanks to the use of only one free parameter in the \citet{Dale2014} model (i.e., the $\alpha$ slope in $dM_{\rm dust} \propto U^{-\alpha}dU$, where $M_{\rm dust}$ and $U$ are dust mass and radiation-field intensity, respectively), the SED-fitting procedure reduces the number of parameters to constrain without far-infrared measurements. Those models are then fitted to the galaxy SEDs with the use of a Bayesian-like analysis. For all models, the $\chi^2$ and the likelihood ($\exp{-\chi^2 / 2}$) for a given observed galaxy are computed, and the value and error of all the parameters returned by CIGALE are the likelihood-weighted mean and standard deviation of all the models, respectively \citep{Boquien2019}.

The quality of the fit is expressed by the reduced $\chi^2$ of the best-fitting model, $\chi_{\rm r}^2$ (i.e., the $\chi^2$ divided by the number of data points). We have applied a quality cut to these fits by removing all objects with a $\chi_{\rm r}^2$ larger than ${\rm median}(\chi_{\rm r}^2)+3$ (in all CIGALE runs considered). This results in 802 objects being removed from the selected sample described in \cref{sec:Sample selection}, before the redshift cuts for D4000$_{\rm w}$/D4000$_{\rm n}$ measurement (the galaxy numbers displayed for these redshift ranges already take into account this quality cut).

For this work, SEDs were fitted to galaxies using: i) PAUS redshift and its corresponding NB photometry (hereafter \textit{PAUS CIGALE}), and ii) CFHTLS redshift \citep{Ilbert2006, Coupon2009} and its broad-band photometry (hereafter \textit{CFHTLS CIGALE}). These two datasets allow us to explore the improvement of the D4000 reconstruction accuracy with narrow-band photometry. As one may expect more improvement fitting both NBs and broad bands, we also run CIGALE with both PAUS NBs and CFHTLS broad-band photometry, but the results were more discrepant with VIPERS spectroscopy than PAUS NBs alone. By adjusting the $\chi^2$  computation procedure \citep{Eriksen2019}, the combination of PAUS and CFHTLS photometry should yield better results than PAUS alone; however, any modifications to CIGALE are left as future work.

\begin{table*}
\centering
	\caption{The input parameters used for SED fitting with CIGALE.}
	\begin{tabular}{l | l}
\hline
Parameter & Values\\
\hline
\multicolumn{2}{c}{Delayed star formation history}\\
\hline
e-folding time of the main stellar population model [Myr] & 100, 300, 500, 1000, 2000, 5000\\
Age [Myr] & 300, 500, 1000, 1500,2500, 4500, 6000\\
\hline 
\multicolumn{2}{c}{Single stellar population \citep{Bruzual2003}}\\
\hline
Initial mass function & \citet{Chabrier2003}\\
Metallicity [$Z_{\odot}$] & 0.02\\
Age of separation between young and old stellar populations [Myr]  & 10\\
\hline
\multicolumn{2}{c}{Dust attenuation law \citep{Charlot2000}}\\
\hline
V-band attenuation ($A_{\rm V}$) in the interstellar medium (ISM) & 0.01, 0.1, 0.4, 0.7, 1.0, 1.5, 2.0, 2.5\\
Power law slopes of the attenuation in the birth clouds (BC) and ISM & -0.7\\
$A_{\rm V}$ ISM / ($A_{\rm V}$ BC +$A_{\rm V}$ ISM) & 0.8\\
\hline
\multicolumn{2}{c}{Dust emission \citep{Dale2014}}\\
\hline
AGN fraction & 0.0, 0.1, 0.3\\
Power law slope dU/dM ($U^{\rm \alpha }$)  & 2.0\\
\hline
\multicolumn{2}{c}{Nebular emission model}\\
\hline
Ionisation parameter & $10^{-2}$\\
Escape fraction of Lyman continuum photons & 0.0\\
Absorption fraction of Lyman continuum photons & 0.0\\
\hline
\end{tabular}
\label{tab:cigale}
\end{table*}

\section{Accuracy and distribution of D4000 estimations with PAUS and CFHTLS data}\label{sec:Accuracy and distribution of D4000 estimations with PAUS and CFHTLS data}

Here, the different methods of D4000 estimation presented in the previous section are compared against each other. The photometric D4000 measurement from PAUS NBs (\cref{sec:Direct estimator}) will simply be referred to as \textit{PAUS direct}, while the D4000 reconstruction with CIGALE (\cref{sec:CIGALE estimation}) will be referred to as explained in the previous subsection.

\subsection{SNR}\label{sec:SNR}

\begin{figure}
 	 \includegraphics[width=\columnwidth]{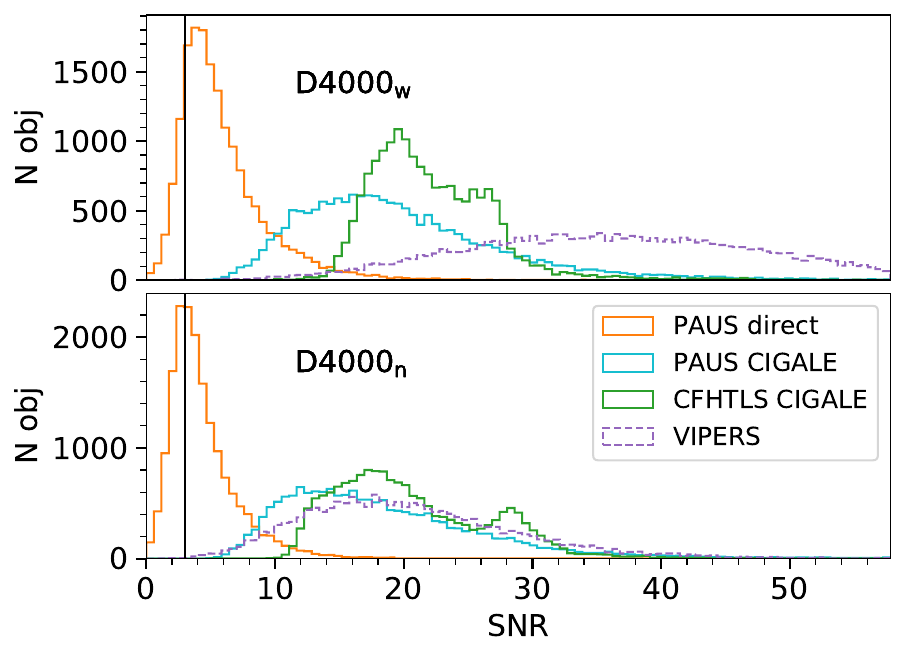}
     \caption{SNR distribution for PAUS direct, PAUS CIGALE, CFHTLS CIGALE, and VIPERS spectroscopy. The black vertical line displays the detection threshold of $\rm SNR > 3$. \textit{Upper panel}: D4000$_{\rm w}$. \textit{Lower panel:} D4000$_{\rm n}$.}
     \label{fig:snr_histogram}
\end{figure}

We have computed both D4000$_{\rm w}$ and D4000$_{\rm n}$ for the sample specified in \cref{sec:Sample selection}, and examined the signal-to-noise ratio (SNR) for each method. \Cref{fig:snr_histogram} displays the SNR distribution for the D4000 estimations with each method (PAUS direct, CIGALE runs and VIPERS spectroscopy), together with the detection threshold we have imposed at $\rm SNR > 3$. This SNR limit has been chosen because it is commonly used in astrophysics as a threshold for source detection in imaging or spectral features, whether it is in the visible spectrum \citep[e.g,][]{Cutri2003, Sanchez2012}, X-ray \citep[e.g.,][]{Bulbul2014} or infrared wavelength range \citep[e.g.,][]{Labbe2013}. The distribution of VIPERS SNR appears as a dashed line; throughout this paper we will use the convention of representing data based on photometric redshifts with solid lines and spectroscopic redshift data with dashed lines.

As shown in \cref{fig:snr_histogram}, the SNR from the CIGALE D4000 reconstruction is significantly larger than the one from PAUS direct. The D4000 values from both CIGALE runs have a mean SNR of $\sim$20, being slightly larger for the CFHTLS case. In both CIGALE runs, all objects in the selected sample are well above the detection threshold, while in the PAUS direct case only 85.01\% (65.87\%) of D4000$_{\rm w}$ (D4000$_{\rm n}$) measurements have $\rm SNR > 3$. The mean SNR from the PAUS direct method is $\sim$5 for the D4000$_{\rm w}$ and $\sim$4 for the D4000$_{\rm n}$. Regarding the spectroscopic VIPERS D4000, we find a mean SNR of $\sim$35 for the D4000$_{\rm w}$ and $\sim$20 for the D4000$_{\rm n}$. The larger SNR in the D4000$_{\rm w}$ for both photometric and spectroscopic measurements is simply due to the larger wavelength range of the D4000 bands for the D4000$_{\rm w}$ definition: if flux is integrated over a larger span, higher SNR is to be expected. The CIGALE D4000 reconstruction, however, seems insensitive to this effect, given that it uses all available flux information to fit a model that predicts a D4000 value. This results in a SNR that is closer to spectroscopy for D4000$_{\rm n}$.

It is reasonable to expect a higher SNR from SED fitting than PAUS direct measurements, given that the spectral information provided to CIGALE spans a much larger wavelength window. CIGALE uses all reference broad bands or PAUS NBs, while in the PAUS direct measurement just the NBs that may contain part of the D4000 bands are taken into account. Moreover, the SED fitting relies on underlying galaxy models, which already places constraints on the possible values of the spectral features, thus constraining the potential dispersion of D4000 values. Nevertheless, these factors alone cannot account for the extreme SNR boost in the CIGALE reconstruction. It is worth noting that CIGALE only uses the observational errors as weights for the $\chi^2$ computation, without propagating them directly into the error budget; the error of the output values (in this case, the D4000) is the likelihood-weighted standard deviation of all fitted models \citep{Boquien2019}. In other words, the noise levels of the photometric fluxes will determine how much these fluxes are weighted in the fit relative to each other, but do not have a direct relationship to the error of the parameter estimated with CIGALE. Therefore, if the PDF of the fit is concentrated in a small probability volume (either because the models do not properly reproduce the diversity of the sample, or because the $\chi^2$ computation tends over-weight specific input data), the error of the output values will be low, even if the input data is noisy. This may result in an error underestimation, and thus explain the high SNR values; the possibility of an error underestimation due to a limited model grid or unaccounted photometric errors is largely discussed in \citet{Noll2009}. However, any modifications or improvements to the SED-fitting code are out of the scope of this work.

\begin{figure}
 	 \includegraphics[width=\columnwidth]{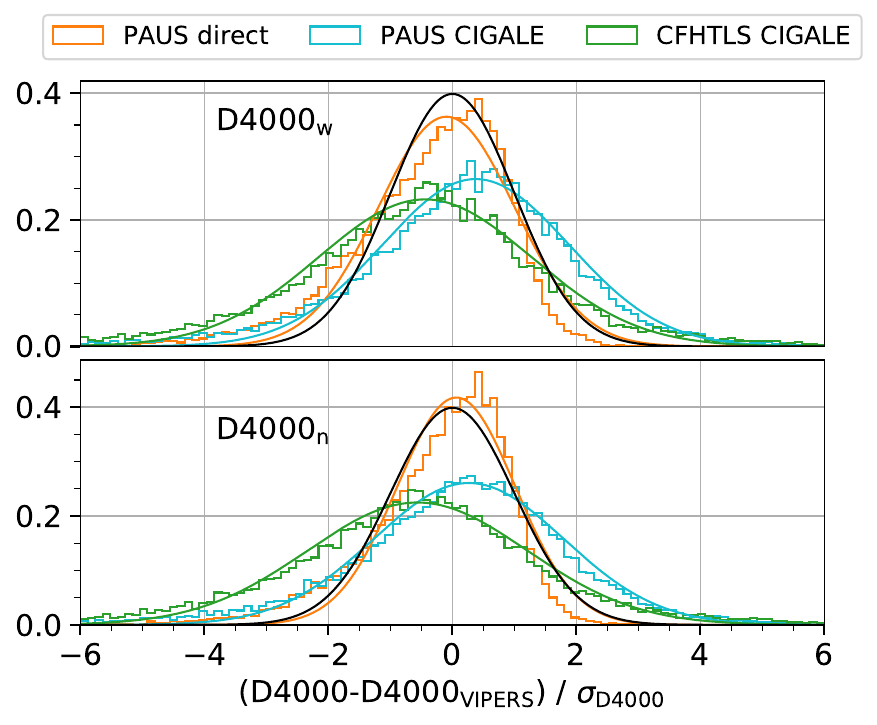}
     \caption{Distribution of the difference between D4000 estimations and VIPERS D4000 values, divided by the estimated error. All distributions are accompanied of their respective Gaussian fit; the solid black line shows a reference Gaussian with $\mu=0$ and $\sigma=1$. \textit{Upper panel}: D4000$_{\rm w}$. \textit{Lower panel:} D4000$_{\rm n}$.}
     \label{fig:error_gaussianity}
\end{figure}

The Gaussianity of the errors associated to these D4000 estimations is evaluated in \cref{fig:error_gaussianity} and \cref{tab:error_fits}. There, we display the distribution of the difference between the D4000 estimations and the true value (i.e., VIPERS spectroscopic measurements), divided by the estimated error. If the errors are properly estimated, these histograms should follow Gaussian distributions of $\mu=0$ and $\sigma=1$. Errors from PAUS direct photometric measurements are realistic, as their histograms in \cref{fig:error_gaussianity} have $\sigma\sim1$ (underestimated by $\sim$10\% for D4000$_{\rm w}$ and overestimated by $\sim$5\% for D4000$_{\rm n}$, see \cref{tab:error_fits}). Both CIGALE methods show clear error underestimations, with $\sigma \sim 1.75$ for CFHTLS CIGALE, and $\sigma \sim 1.5$ for PAUS CIGALE, which is akin to a error underestimation of 75\% and 50\% respectively. Hence, the SNR values derived from the CIGALE reconstructions in \cref{fig:snr_histogram} should be divided at least by a factor equal to their $\sigma$ in \cref{tab:error_fits} to be considered realistic. Despite the PAUS direct D4000 measurement providing a realistic error estimate consistent with a Gaussian behaviour, and the CIGALE D4000 reconstruction overestimating its SNR, it is worth noting that by construction, CIGALE still provides less dispersion. Due to the limited grid of SED models, the reconstructed CIGALE D4000 will always have a realistic value within the range allowed by the models, regardless of how faint the observed object is.

\begin{table}
\centering
	\caption{Parameters of the Gaussian fits to all error distributions displayed in \cref{fig:error_gaussianity}.}
 	\label{tab:error_fits}
 	\begin{tabular}{rrcc}
 		\hline
 		 & & $\mu$ & $\sigma$\\
 		\hline
 		\multirow{2}{*}{PAUS direct} & D4000$_{\rm w}$ & $-0.10\pm0.03$ & $1.10\pm0.02$ \\
 	    & D4000$_{\rm n}$ & $0.06\pm0.02$ & $0.96\pm0.02$\\
 	    \hline   		
 	    \multirow{2}{*}{PAUS CIGALE} & D4000$_{\rm w}$ & $0.38\pm0.02$ & $1.51\pm0.02$ \\
 	    & D4000$_{\rm n}$ & $0.27\pm0.02$ & $1.53\pm0.02$\\
		\hline
  		\multirow{2}{*}{CFHTLS CIGALE} & D4000$_{\rm w}$ & $-0.44\pm0.02$ & $1.72\pm0.02$ \\
 	    & D4000$_{\rm n}$ & $-0.57\pm0.02$ & $1.77\pm0.02$\\
 	    \hline
    \end{tabular}
\end{table}

\begin{figure}
 	 \includegraphics[width=\columnwidth]{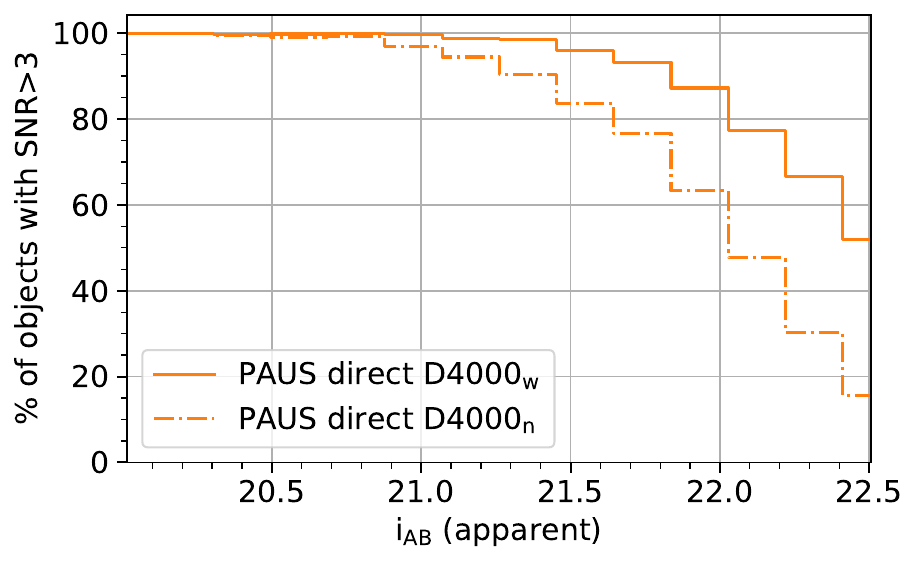}
     \caption{Percentage of objects with the photometric D4000 measurements with PAUS direct $\rm SNR > 3$, in bins of apparent magnitude. Solid line represents D4000$_{\rm w}$, dash-dotted line D4000$_{\rm n}$.}
     \label{fig:snr_vs_magnitude}
\end{figure}

Given the significant percentage of objects with $\rm SNR<3$ in PAUS direct, we have evaluated up to which magnitude PAUS direct allows an individual photometric D4000 measurement of all galaxies above the established SNR threshold ($\rm SNR>3$). In \cref{fig:snr_vs_magnitude}, the percentage of objects with $\rm SNR > 3$ in apparent magnitude bins is shown. Both photometric D4000$_{\rm w}$ and D4000$_{\rm n}$ have $\rm SNR>3$ for almost all objects up to $i_{\rm AB}<21$ (99.88\% and 98.97\% for D4000$_{\rm w}$ and D4000$_{\rm n}$ respectively), with the percentage of objects with measurable D4000 decreasing above this threshold. Consequently, we define a $\rm SNR>3$ D4000 subsample by applying an additional $i_{\rm AB}<21$ cut to the sample defined in \cref{sec:Sample selection}; this leaves a total of 2,534 galaxies. We will refer to this $i_{\rm AB<21}$ sample as \textit{bright sample} henceforth, while the original sample defined in \cref{sec:Sample selection} will be referred to as \textit{full sample}. We will use this bright sample to evaluate how well PAUS direct performs when all objects have a SNR acceptable for individual measurements, while the full sample will be employed to evaluate how well can we recover mean sample trends despite the low SNR.

The advantage of this bright sample is that it preserves PAUS magnitude completeness (which large spectroscopic surveys do not achieve, due to target selection) while ensuring that the PAUS direct individual photometric measurements have high enough SNR. For the bright sample, the average PAUS direct SNR is 11.42 for D4000$_{\rm w}$ and 8.63 for D4000$_{\rm n}$. At $i_{\rm AB}=23$, the magnitude limit for the PAUS catalogue, only $\sim$30\% of the objects have a D4000$_{\rm w}$ above detection threshold for the photometric measurements, while for D4000$_{\rm n}$ this percentage decreases to just $\sim$12\%. Hence, with the narrow-band configuration of PAUS we can achieve D4000 individual measurements with $\rm SNR > 3$ for a magnitude-completeness threshold two magnitudes brighter than the magnitude limit of the full catalogue. It is reasonable to expect similar performance with comparable narrow-band surveys, such as J-PAS \citep{Benitez2014}.

\subsection{Distribution and bias}\label{sec:Distribution and bias}

\begin{figure*}
 	 \includegraphics[width=\textwidth]{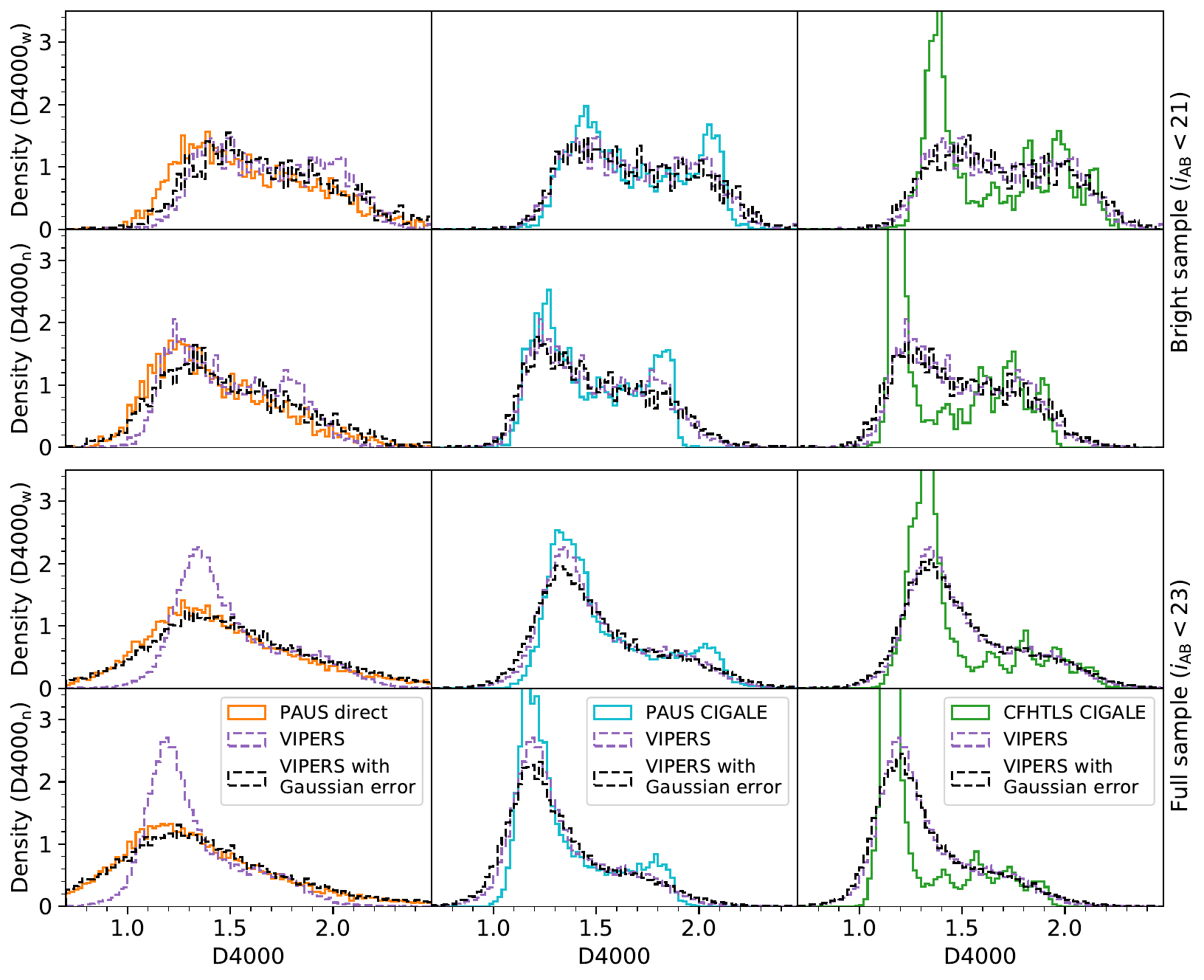}
     \caption{Histogram of the D4000$_{\rm w}$ (even rows) and the D4000$_{\rm n}$ (odd rows) for the photometric PAUS direct measurement (left column), PAUS CIGALE reconstruction (middle column), and CFHTLS CIGALE reconstruction (right column). The two upper rows correspond to the bright sample, while the two lower rows display the full sample. All panels include the distribution of the spectroscopic VIPERS D4000 measurement (purple dashed line), as well as the distribution of the test dataset generated by adding a Gaussian value of mean zero and $\sigma$ equal to its respective non-spectroscopic error to the VIPERS D4000 values (black dashed line).}
     \label{fig:D4000_hist}
\end{figure*}

Once we have examined the SNR and error of the D4000 estimation methods, we study the distribution of the D4000 values themselves. \Cref{fig:D4000_hist} shows a histogram of D4000 values for each method, and for both the bright and full samples. In addition to this, we have also generated test datasets for PAUS direct, PAUS CIGALE and CFHTLS CIGALE D4000 by taking the spectroscopic VIPERS D4000 measurements for each object, and adding to them a value drawn from a Gaussian distribution of mean zero and $\sigma$ equal to the error of its respective non-spectroscopic estimation. For the CIGALE D4000 reconstructions, we use the error provided by CIGALE as it is, without any correction to mitigate the error underestimation shown in \cref{fig:error_gaussianity} and \cref{tab:error_fits}.

If the D4000 estimations from these non-spectroscopic methods are unbiased and with an error estimation that behaves like a Gaussian, the D4000 distribution from the test dataset generated by adding these artificial errors to VIPERS D4000 should be the same as the D4000 distribution from its respective non-spectroscopic method. We know from \cref{fig:error_gaussianity} that the error in the photometric PAUS direct measurement is reasonably close to Gaussian, and that for the CIGALE reconstructions it is clearly underestimated; these test datasets allow us to check if the differences between the VIPERS D4000 distribution and the non-spectroscopic D4000 distributions can be accounted for solely by the non-spectroscopic error. The test datasets are represented by dashed black lines.

The spectroscopic D4000 distributions in \cref{fig:D4000_hist} (purple dashed line) exhibit a weak bimodality, with a peak centred on the blue cloud around 1.3 and 1.2 for D4000$_{\rm w}$ and D4000$_{\rm n}$, respectively. This blue peak is far stronger for the full sample: as blue galaxies are fainter on average, by imposing the $i_{\rm AB}<21$ cut most of the removed galaxies belong to the blue cloud. On the other hand, the red sequence appears as a plateau in the much longer tail of the high D4000 end. The bimodality in the D4000 distribution has been extensively demonstrated in the literature \citep[see e.g.,][]{Kauffmann2003b, Haines2017} and it has been shown to become less pronounced as redshift increases \citep[e.g.,][fig. 1]{Haines2017}. Therefore, the weak bimodality in our sample at $z>0.5$ is to be expected.
 
In order to quantify the agreement between the D4000 distribution of the different methods and the VIPERS D4000, we have performed the Epps-Singleton (ES) test \citep{Epps1986} on all cases displayed in \cref{fig:D4000_hist}, comparing the D4000 distributions against the noiseless spectroscopic VIPERS D4000 measurements, and the VIPERS test datasets generated with the non-spectroscopic noise (black lines in \cref{fig:D4000_hist}). The ES test has been chosen instead of other popular alternatives such as the Kolmogorov-Smirnov \citep{Massey1951}, Cramer-von Mises \citep{Anderson1962} or Anderson-Darling \citep{Anderson1952} tests because it was the least dependant on the D4000 range where the empirical distribution function (EDF) was determined. This result is expected, given that the ES test has been specifically developed to provide enough power under a great variability of EDF parameters, such as location, scale or even family of distributions \citep[see][]{Epps1986}, and in our case the tails of the D4000 distribution greatly vary between D4000 methods. For the PAUS direct measurement, the tails are significantly larger than VIPERS due to the lower SNR, and for both CIGALE reconstructions the tails end abruptly given the limitations of the SED models (see \cref{fig:D4000_hist}).

The D4000 EDF has been computed in all cases with a uniform binning between $\rm D4000=0.7$ and $\rm D4000=2.5$, with a bin width $\rm \Delta D4000 = 0.02$; this is the same binning as the histograms displayed in \cref{fig:D4000_hist}. The \textit{p}-values of the ES tests are shown in \cref{tab:ES_test_results}; these are the probabilities of the estimated D4000 values coming from the same distribution as the noiseless spectroscopic VIPERS D4000 (or VIPERS D4000 with non-spectroscopic noise, in parentheses).

\begin{table}
\centering
	\caption{\textit{p}-values of the ES test performed for all the D4000 distributions displayed in \cref{fig:D4000_hist}. For each case, the \textit{p}-values from the comparison of the respective D4000 distribution to VIPERS D4000 are shown first. The \textit{p}-values of the comparison to VIPERS D4000 with the respective non-spectroscopic Gaussian noise (black lines in \cref{fig:D4000_hist}) are displayed in parentheses.}
 	\label{tab:ES_test_results}
 	\begin{tabular}{rccc}
 		\hline
   		 & PAUS direct  & PAUS CIGALE & CFHTLS CIGALE\\
 		\hline
 		\multicolumn{4}{l}{Bright sample ($i_{\rm AB}<21$)}\\
 		\hline
        D4000$_{\rm w}$ & 0.50 (0.93) & 0.81 (0.60) & 0.25 (0.09)\\
        D4000$_{\rm n}$ & 0.19 (0.88) & 0.71 (0.35) & 0.10 (0.03)\\
        \hline
 		\multicolumn{4}{l}{Full sample ($i_{\rm AB}<23$)}\\
 		\hline
        D4000$_{\rm w}$ & 0.02 (0.98) & 0.84 (0.58) & 0.55 (0.38)\\
        D4000$_{\rm n}$ & 0.00 (0.99) & 0.52 (0.20) & 0.17 (0.09)\\ 		
        \hline
 	\end{tabular}
\end{table}
 
The distribution of the PAUS direct D4000 (left column in \cref{fig:D4000_hist}) shows the weakest bimodality of all the non-spectroscopic methods, with a much smaller peak in the blue cloud, and a less pronounced difference between the high and low D4000 tails. This effect is caused by the noise in the photometric D4000 measurements, and is far more noticeable for the full sample due to the much lower average SNR. In fact, the \textit{p}-values for the bright sample when compared to spectroscopic VIPERS measurements are far from negligible (up to 0.5 in D4000$_{\rm w}$, see \cref{tab:ES_test_results}) while for the full sample they tend to zero. However, for the PAUS direct measurements the black line representing the spectroscopic VIPERS D4000 with PAUS direct Gaussian errors shows an excellent agreement (\textit{p}-$\rm value \sim0.9$ and \textit{p}-$\rm value \sim1$ for the bright and full samples respectively), with the only noticeable difference being a shift of the distribution peak of $\sim$0.1 towards lower values. This is consistent to a certain extent with the negative bias of the photometric PAUS direct estimator shown in \cref{sec:Synthetic PAUS tests}. Nevertheless, this similarity between distributions is a clear indicator that the error in PAUS direct is close to Gaussian.

Regarding the PAUS CIGALE reconstruction, we can see in \cref{fig:D4000_hist} (centre column) that the D4000 distribution is much less affected by noise than in the PAUS direct case. This difference is far more noticeable in the full sample, where the PAUS direct distribution is heavily smoothed by noise, but the PAUS CIGALE distribution follows the VIPERS D4000 distribution far more closely (\textit{p}-$\rm value =0.84$ and \textit{p}-$\rm value =0.52$ for D4000$_{\rm w}$ and D4000$_{\rm n}$, respectively). When we add the CIGALE D4000 reconstruction error to the VIPERS D4000 distribution, the differences between the CIGALE and VIPERS D4000 distributions increase (a drop of $\sim$0.3 in \textit{p}-values), since the Gaussian error further smooths the VIPERS D4000 distribution, but the PAUS CIGALE distributions are actually "sharper". This "sharpening" of the PAUS CIGALE distributions is an artefact of the SED-fitting methodology; the reconstructed D4000 values tend to cluster around the underlying SED templates, resulting in stronger peaks around the most representative red/blue galaxies than in the real sample (i.e., \textit{artificial bimodality}). The artificial bimodality is far less evident with the D4000$_{\rm w}$ reconstruction, which may be due to the smaller difference in D4000 values between SED templates due to the larger continuum range, or the larger dependency on dust extinction smoothing this difference. In the case of the CFHTLS CIGALE distribution (\cref{fig:D4000_hist}, right column), the artificial bimodality is far more exaggerated, thus resulting in a much larger discrepancy with spectroscopy (lower \textit{p}-values all across \cref{tab:ES_test_results}).

\begin{figure*}
 	 \includegraphics[width=\textwidth]{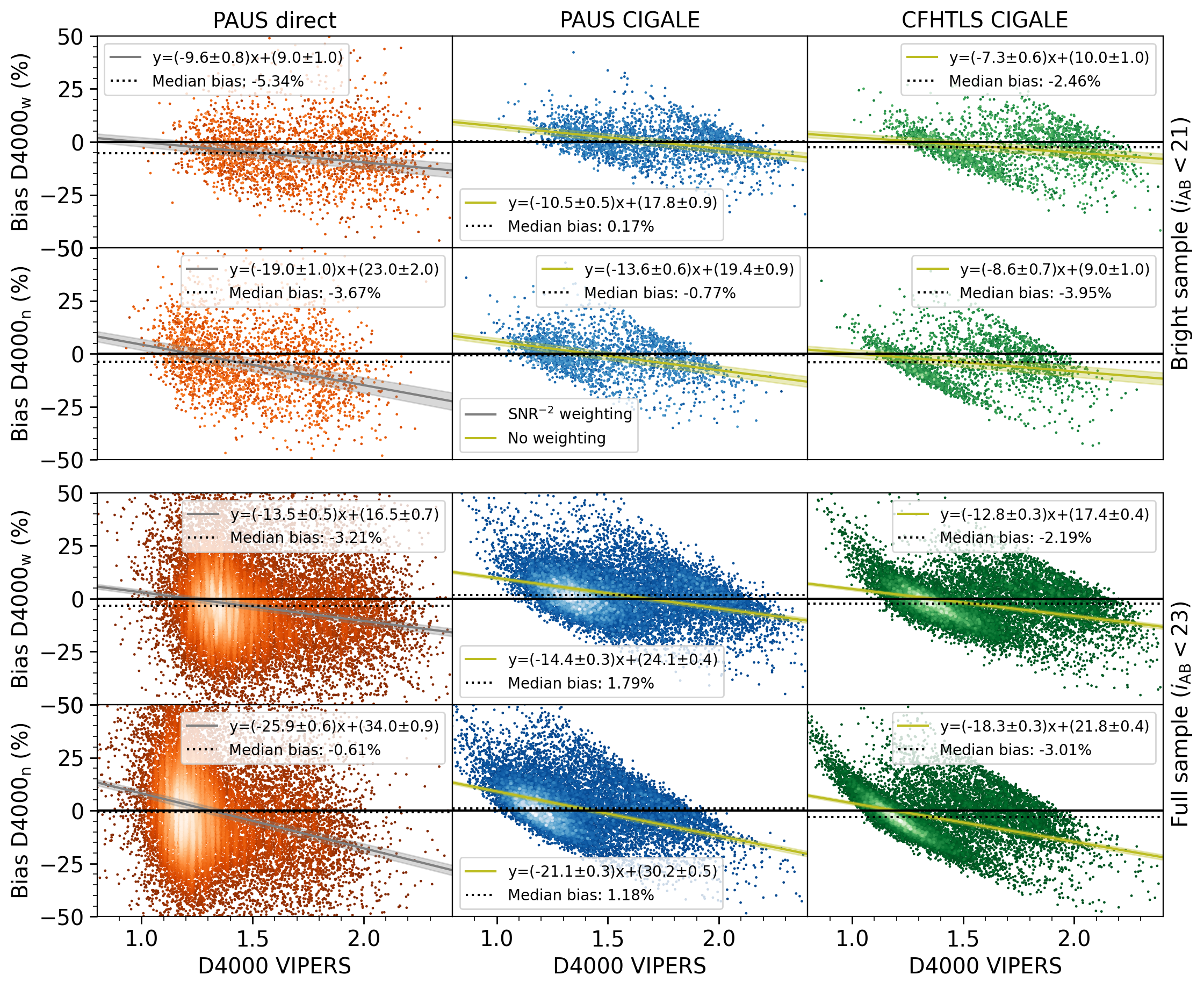}
     \caption{Bias in the D4000 measurements for all non-spectroscopic D4000 estimation methods, considering the bright and full sample. The arrangement of the panels is analogous to \cref{fig:D4000_hist}. In each panel, the x-axis values are the VIPERS counterpart of the respective D4000 definition. For the scatter plots, the colour lightness is proportional to the point density, with lighter areas being denser. A linear fit to the bias is represented as a grey or yellow line, depending on the weighting used for the fit (SNR$^2$ or no weighting, respectively). In all panels, the dotted black line represents the median bias of the sample. Legends show the value of this median bias, as well as the parameters of the linear fit.}
     \label{fig:scatter_bias}
\end{figure*}

In addition to the D4000 distributions, we also study the bias of each D4000 estimation method versus the spectroscopic D4000 value (see \cref{fig:scatter_bias}, which is the observational counterpart of \cref{fig:bias_scatter_synth}). In \cref{fig:scatter_bias}, each panel displays a scatter plot with the VIPERS D4000 measurements in the x-axis, and the bias of the respective D4000 estimation method in the y-axis. This bias is determined with \cref{eq:bias_definition}; but this time with the D4000 estimated from actual PAUS observations (either with direct photometric measurements or CIGALE reconstructions). Both the bright and full sample are displayed, following the same panel arrangement as in \cref{fig:D4000_hist}. Overall, the biases present a much larger scatter in \cref{fig:scatter_bias} than in \cref{fig:bias_scatter_synth}, especially in the full sample, but this is due simply to the photometric noise, which simply does not exist in \cref{fig:bias_scatter_synth}. Hence, our analysis in this subsection will be focused on the median bias and the average linear correlation between bias and the spectroscopic D4000 value.

We have performed linear fits to all datasets to evaluate the average correlation between the bias and the VIPERS D4000; all fits are unweighted except the PAUS direct measurement, where a SNR$^2$ weighting has been applied. The reason for this weighting is that the high noise of the PAUS direct D4000 resulted in a linear fit with large error regions, which provided close to no meaningful information about the actual trend. A SNR$^2$ weighting has been chosen instead of $\sigma^{-2}$ to mitigate the dependence of the error with the D4000 value: when weighting objects with different D4000, the SNR is a more accurate estimator of the amount of information per measurement. We have tested this weighting by repeating the SNR$^2$-weighted linear fit on the CIGALE D4000 reconstructions, but using PAUS direct measurement SNR: the resulting linear fit was compatible with its unweighted counterpart (the error regions of the fits fully overlapped). Hence, we can affirm that the SNR$^2$ weighting does not significantly distort the correlation between bias and VIPERS D4000. It is worth noting that such weighting can not be performed with the CIGALE errors, as there is a far more significant dependence of errors vs D4000 (almost an order of magnitude, with errors increasing with D4000 value, reaching a maximum around $\rm D4000\sim1.8$). This error-D4000 correlation in the CIGALE D4000 reconstructions distorts the trend found in the unweighted linear fit.

The PAUS direct measurement (left column) displays a mean negative bias of the order of few percent, with the linear fit showing a negative correlation between the D4000 value and bias. This negative correlation comes from the low spectral resolution of the photometric data used. Since PAUS NBs have a FWHM of the same order of magnitude as the D4000 bands, the red D4000 band may contain some flux information from the slope of the spectral break itself, which would bias its measured flux towards higher values. The opposite may happen with the blue D4000 band, where including flux from the slope of the spectral break will decrease the measured flux (\cref{fig:D4000_measurement_example} may help better visualising this effect). The result of these two effects is an underestimation of the D4000 that becomes more pronounced as the D4000 value increases (see \cref{eq:D4000_definition}), a trend consistent with the linear fit to the bias displayed in \cref{fig:scatter_bias}.

Following with the PAUS direct comparison, the D4000$_{\rm w}$ measurement shows a weaker bias correlation than D4000$_{\rm n}$ (with the slope of the linear fit being approximately half as steep), but a larger absolute value of the median bias (by 2-3\%). Given that a fixed bias offset is far simpler to account for and correct than a linear trend depending on the true D4000 values, we consider the PAUS direct measurement of the D4000$_{\rm w}$ to behave better than the D4000$_{\rm n}$. When comparing the full sample against the bright sample, we see that the bias slope is slightly steeper, and that the absolute value of the median bias is actually smaller by 2-3\%. Both of these effects can be accounted for by the low SNR of faint objects in the PAUS sample, given that the D4000 is the ratio of two fluxes. If the flux in the denominator (the blue band, see \cref{eq:D4000_definition}) is noisy enough to be close to zero, it will result in an unrealistically high D4000 value (albeit with very low SNR). If we apply a very simple cut to these high outliers (i.e., PAUS direct D4000 < 5), both the median bias and the slope of the full sample become significantly closer to the bright sample results.

Regarding the PAUS CIGALE D4000 reconstruction (centre column), the median bias has an absolute value of the order of $\sim$1\%; however, it is not strictly negative as in PAUS direct. There also appears a negative correlation between VIPERS D4000 and bias, with a slope roughly similar to PAUS direct (slightly steeper for D4000$_{\rm w}$, and a bit less pronounced for D4000$_{\rm n}$). In this case, the negative slope is mostly caused by the artificial limits that the SED fitting imposes on the D4000 values; it can be clearly appreciated in the centre column of \cref{fig:scatter_bias} that the distribution of the data points are clearly cut by two diagonal lines at the low and high D4000 ends. The objects with high VIPERS D4000 values display a strong negative bias (e.g., at VIPERS D4000$_{\rm n}>1.9$), and the objects with low VIPERS D4000 values have a pronounced positive bias (e.g., at VIPERS D4000$_{\rm n}<1.0$). 

Comparing the CFHTLS CIGALE results (left column) to PAUS CFHLTS, we see that for CFHTLS the correlation slopes are somewhat less pronounced, but there is a negative median bias of the order of -2-3\%, similarly to PAUS direct. Moreover, there is a strong clustering of data points in a stripe pattern, due to the tendency of the SED-fitting methodology to infer specific D4000 values. One possible explanation for these stripes is the finite number of SED templates in the SED-fitting code, with the D4000 reconstructions tending to cluster around the most representative SEDs for red and blue galaxies (explaining also the stronger bimodality). This stripe pattern is much less noticeable in PAUS CFHTLS, which is in accordance with the smoother D4000 distribution in \cref{fig:D4000_hist}.

Therefore, we can conclude from this section that the PAUS direct D4000 measurement is fully consistent with the VIPERS D4000 plus a Gaussian noise, which is properly estimated by simple propagation of the photometric errors. In the bright sample (where PAUS direct D4000 $\rm SNR > 3$), the performance is similar to CIGALE SED fitting with PAUS photometry, with the advantage of being completely model-independent (and requiring negligible computational cost). In the full sample, where a significant fraction of the PAUS direct D4000 measurements have $\rm SNR < 3$, the results can be kept consistent if measures are taken to account for this noise (i.e., simulating the noise in the VIPERS D4000 in \cref{fig:D4000_hist}, or using SNR$^2$ or median-based statistics in \cref{fig:scatter_bias}). Regarding the CIGALE results, we see that PAUS CIGALE performs equally well in both the bright and full samples, and significanly outclasses the CFHTLS CIGALE reconstruction both in error estimation (\cref{tab:error_fits}) and D4000 distribution (\cref{fig:D4000_hist}), while providing arguably less biased measurements (\cref{fig:scatter_bias}). This is a clear proof of narrow-band photometry providing a net benefit to SED-fitting analysis.

Henceforth, we will not take into account the CFHTLS CIGALE D4000 in our comparisons, as we consider proven that the PAUS CIGALE D4000 outperforms it. Moreover, when we evaluate the evolution of average sample trends, we will use the full sample, as the low SNR of the PAUS direct D4000 should not be an impediment to retrieve results compatible with spectroscopy. On the other hand, when we analyse properties related to individual galaxies (e.g., galaxy classification), we will consider both the bright and full samples, in order to evaluate how well the PAUS direct photometric measurement performs in a sample where all objects have $\rm SNR > 3$.

\section{D4000 and galaxy properties}\label{sec:D4000 and galaxy properties}

Here, we examine how the D4000 estimations vary across different observational parameters. For all the results displayed in this section, we have applied additional cuts to our galaxy sample: all objects with $\rm SNR<3$ in the VIPERS D4000 (either D4000$_{\rm w}$ or D4000$_{\rm n}$) have been removed, as well as spectroscopic outliers. We have considered outliers all objects outside of the range $0.7<\rm VIPERS\, D4000<2.5$, again, for both D4000 definitions. These cuts removed an extra 46 objects from the sample defined in \cref{sec:Sample selection}, resulting in 17,195 galaxies.

\subsection{D4000 versus redshift}\label{sec:D4000 versus redshift}

\begin{figure*}
 	 \includegraphics[width=\textwidth]{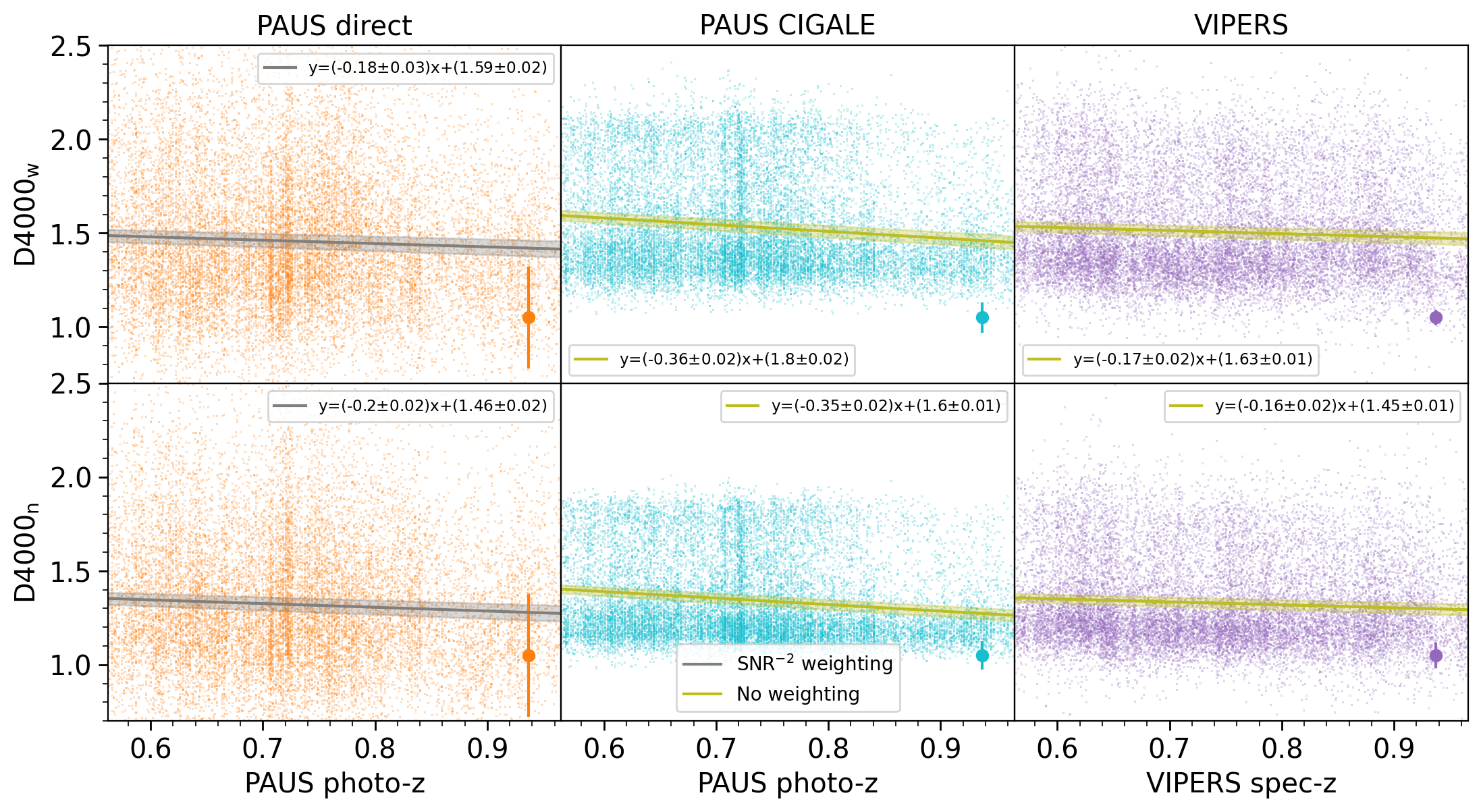}
     \caption{D4000$_{\rm w}$ (upper row) and D4000$_{\rm n}$ (lower row) versus redshift for PAUS direct measurement (left column), PAUS CIGALE reconstruction (centre column), and VIPERS D4000 measurement (right column). PAUS photo-z used in left and centre columns, and VIPERS spec-z in the right column. The linear fit to each dataset is displayed as a grey or yellow line, depending on the weighting used for the fit (SNR$^2$ or no weighting, respectively). The numerical values of the fit are shown in the legend of each plot. The dot with errorbars in the lower-right corner of each plot represents the median error in the respective D4000 estimation.}
     \label{fig:D4000_vs_z}
\end{figure*}

The first parameter we have considered is redshift: \cref{fig:D4000_vs_z} displays scatter plots of the different D4000 measurements versus redshift, for the full sample. PAUS data (PAUS direct measurements and PAUS CIGALE D4000 reconstructions) have PAUS photo-z in the x-axis, and for VIPERS D4000 measurements its respective spec-z is used. In the panels using PAUS photo-z a sharp stripe of data points appears around $z\sim0.7$, but is absent from VIPERS spec-z. This is due to the large horizontal stripes in \cref{fig:z_comparison} around that same redshift. While this is an artefact of the current photo-z computations, it does not affect the average D4000 properties, as it is seen throughout this section. Moreover, such artefacts should be mitigated in future data releases.

In order to quantify the D4000 evolution with redshift, we perform linear fits to the data points, using a SNR$^2$ weighting for PAUS direct and no weighting for PAUS CIGALE and VIPERS (following the reasoning described in \cref{fig:scatter_bias}). We have confirmed that the SNR$^2$ weighting does not distort the trend by applying the same weighting (with PAUS direct SNR) to the VIPERS D4000, and retrieving linear fits compatible with the unweighted results. For all cases, we find a decreasing trend of the D4000 with redshift: for the PAUS direct measurement, this trend is in agreement within 1$\sigma$ with spectroscopy. However, for PAUS CIGALE the slope of the linear fit is approximately twice as steep as its spectroscopic counterpart (despite PAUS CIGALE mimicking better the true VIPERS D4000 distribution). The reason for this exaggerated trend in the PAUS CIGALE reconstruction might lie in the smaller wavelength range of PAUS NBs compared to CFHTLS broad bands: at a given redshift, the wavelength window to perform the SED fitting is smaller, even if the spectral resolution is higher. This may result especially problematic in the edges of our redshift range, where the D4000 bands are close to the wavelength limits. Moreover, undesired biases at these redshift limits would affect more the slope of the linear fit than at intermediate redshift values.

At higher redshift stellar populations are younger, and thus a decrease of the average D4000 is to be expected, given that it is a proxy for stellar ages. Despite the selection effects at different redshifts of a magnitude-limited sample, this decreasing trend is noticeable in \cref{fig:D4000_vs_z}. This trend is also widely documented in the literature. In fact, with the same VIPERS data used in this work \citep{Scodeggio2018}, a decreasing trend of D4000$_{\rm n}$ versus redshift can be seen in \citet{Siudek2017}, fig. 7, and \citet{Haines2017}, table 1 \citep[far more examples can be found in literature at different redshifts, e.g.][]{Moresco2012, Joshi2019, Borghi2021}.

\subsection{Galaxy classification}\label{sec:Galaxy classification}

The classification of galaxies into different categories has been one of the main topics of galaxy study ever since the very beginnings of observational cosmology, with the first example being the morphological classification into \textit{early-type} (elliptical) and \textit{late-type} (spiral) galaxies \citep{Hubble1926, DeVaucouleurs1959}. More sophisticated morphological classifications have been developed \citep[e.g.,][]{Strateva2001, Driver2006, Mignoli2009, Krywult2017}, as well as classifications based on other observables: colours \citep[e.g.,][]{Arnouts2013, Taylor2015}, spectral features \citep[e.g.,][]{Kauffmann2003a}, or a combination of them \citep[e.g.,][]{Siudek2018, Turner2021}, including the D4000. It is natural for the simplest classification models to separate galaxies into two different types that mimic the bimodality in their parameters. In this subsection, we will evaluate how well a D4000-based classification with the accuracy of the different methods can separate the blue and red galaxy populations, as defined by a fiducial classification.

The fiducial classification used in this work is presented in \citet{Siudek2018}; it has been carried out over the entire VIPERS data release \citep{Scodeggio2018} with a Fisher Expectation-Maximisation unsupervised algorithm, using as entry parameters 12 different rest-frame magnitudes (normalised to $i_{\rm AB}$) and spectroscopic redshift. The algorithm have distinguished 11 different galaxy classes, with an extra class of broad-line AGNs that has been omitted for this work. These classes are defined in order: class 1 corresponds to the reddest galaxies, and class 11 to the bluest ones. For the binary red/blue classification used in this work, we have considered all objects of class 6 and below to be red, and 7 and above to be blue.

\begin{figure}
 	 \includegraphics[width=\columnwidth]{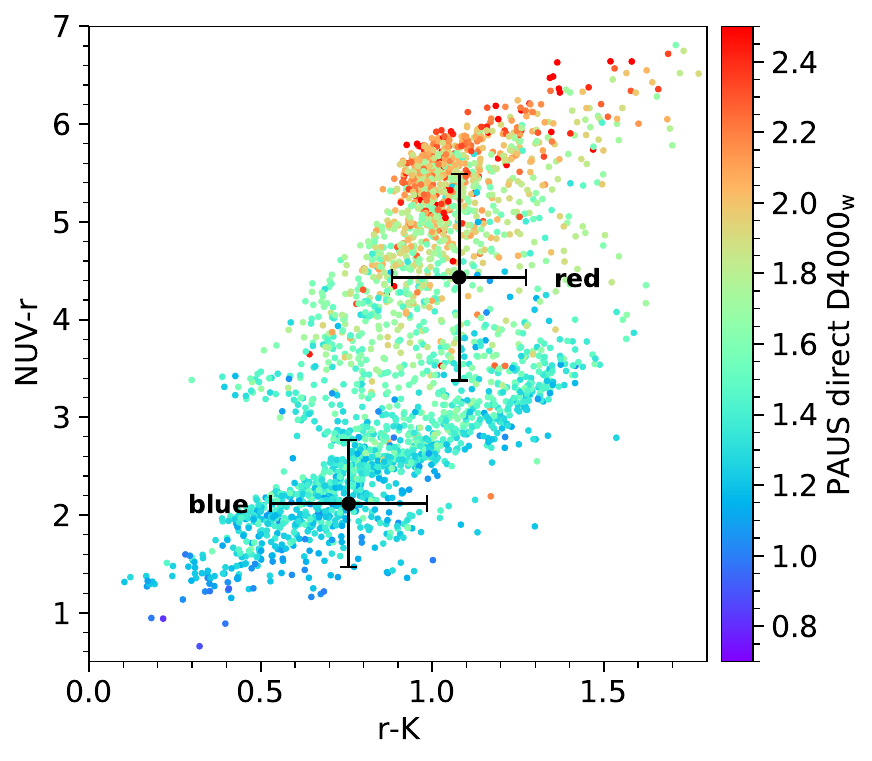}
     \caption{$NUVrK$ colour diagram for the bright sample, colour-coded with the PAUS direct D4000$_{\rm w}$ (with colour scale limited by the outlier cuts defined for VIPERS D4000). The average positions of the red/blue galaxies from our fiducial classification are also displayed, with their errorbars being their standard deviations.}
     \label{fig:scatter_classes}
\end{figure}

This separation between red/blue galaxies is also visible on the $NUV-r$ vs $r-K$ diagram \citep[$NUVrK$][]{Arnouts2013, Moutard2018}. In \cref{fig:scatter_classes}, we show the $NUVrK$ colour diagram for the bright sample ($i_{\rm AB}<21$), with the data points colour-coded with their PAUS direct D4000$_{\rm w}$ measurements, and the $NUV-r$ and $r-K$ colours being extracted from the PAUS CIGALE run. The average positions in the diagram of the PAUS red and blue galaxies, according to our fiducial classification, are also displayed. and are coherent with the aforementioned references, as well as with the average positions of galaxy classes in the $NUVrK$ colour diagrams \citep[see fig. 1 in][]{Siudek2018}.

This fiducial galaxy classification is used to test the performance of a single D4000 cut (hereafter D4000$_{\rm cut}$) in order to distinguish between red and blue galaxies, with a galaxy being blue if $\rm D4000 < D4000_{\rm cut}$, and red otherwise. For each D4000 estimation method, we have tested D4000$_{\rm cut}$ values in a physically meaningful range that comprises all but the reddest/bluest galaxies ($1.1<{\rm D4000}_{\rm cut}<1.9$), and determined the percentage of galaxies correctly classified as red/blue with each given cut. This D4000$_{\rm cut}$ range has been chosen because it is unreasonable to evaluate D4000$_{\rm cut}=1$ or D4000$_{\rm cut}=2$ as thresholds to separate between red and blue galaxies, since these values correspond to highly star-forming and very quenched galaxies, respectively.

\begin{figure}
 	 \includegraphics[width=\columnwidth]{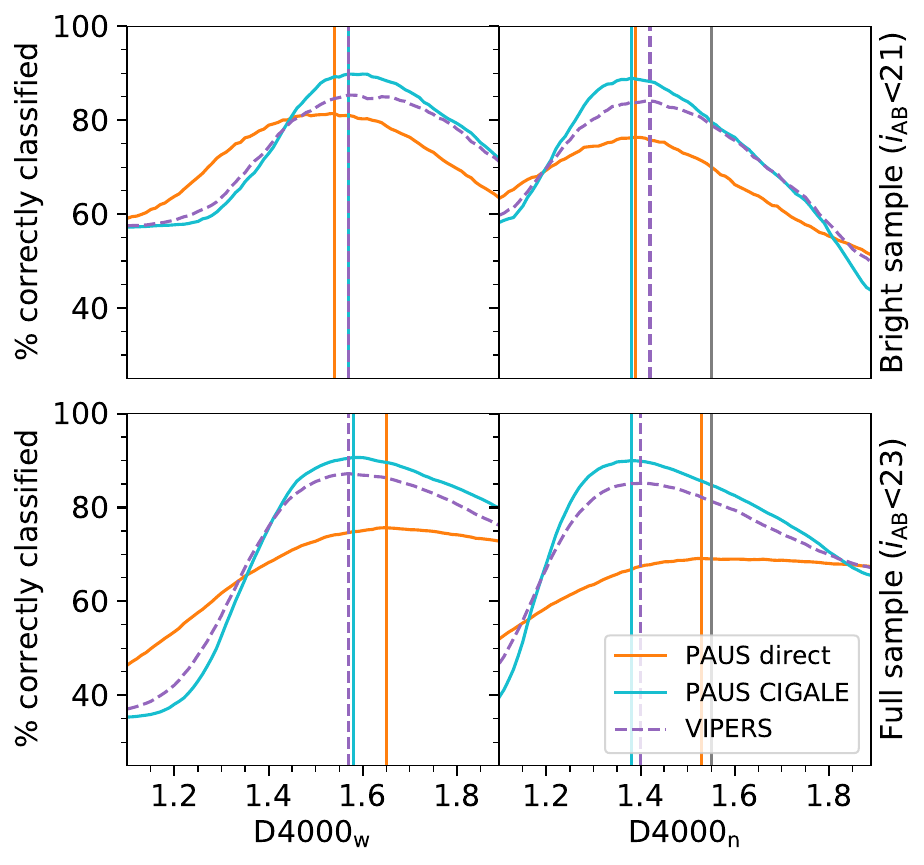}
     \caption{Percentage of correctly classified objects in the red/blue galaxy sample determined with a given D4000$_{\rm cut}$, using as a fiducial galaxy classification the results of \citet{Siudek2018}. Vertical lines represent the D4000 threshold for maximum percentage for each D4000 estimation method (orange for PAUS direct measurements, cyan for PAUS CIGALE reconstruction and purple for VIPERS spectroscopic measurements). Left column shows results for D4000$_{\rm w}$, right column for D4000$_{\rm n}$. Upper row displays the results only for the bright sample ($i_{\rm AB}<21$), while lower row for the full sample. The standard D4000$_{\rm cut\, n}=1.55$ is also shown as a vertical grey line in the right column.}
     \label{fig:D4000_threshold}
\end{figure}

The results are given in \cref{fig:D4000_threshold} and \cref{tab:D4000_cut_results}. In order to check the consistency of these results with the literature, we also evaluate the percentage of correctly classified galaxies using the cut of D4000$_{\rm cut\, n}=1.55$ as defined in \citet{Kauffmann2003b} \citep[and other works, e.g.,][]{Haines2017}. Since this cut is defined for the D4000$_{\rm n}$, we do not perform this comparison for the D4000$_{\rm w}$. This D4000$_{\rm cut\, n}=1.55$ value is significantly higher than the D4000$_{\rm cut\, n}=1.4$ we find for the VIPERS D4000 measurements in the full sample. This discrepancy may simply arise from the different classification criteria adopted, but it is remarkable that the difference between the D4000$_{\rm cut \, n}$ derived from the fiducial classification \citep{Siudek2018} and the D4000$_{\rm cut\, n}=1.55$ proposed in \citet{Kauffmann2003b} is smaller than the difference between the D4000$_{\rm n}$ and D4000$_{\rm w}$ cuts derived from the fiducial classification alone. Hence, the D4000 definition may play a larger role in the value of the optimal value of the D4000 cut than the classification criteria themselves.

\begin{table*}
\centering
	\caption{D4000$_{\rm cut}$ that ensures the maximum percentage of correctly classified red/blue galaxies, for all cases represented in \cref{fig:D4000_threshold}. The percentage of correctly classified galaxies with this optimal D4000$_{\rm cut}$ is displayed in parentheses, while the same percentage for the standard D4000$_{\rm  cut\, n}=1.55$ is shown in brackets (only for D4000$_{\rm n}$).}
 	\label{tab:D4000_cut_results}
 	\begin{tabular}{rccc}
 		\hline
 		 & PAUS direct  & PAUS CIGALE & VIPERS\\
 		\hline
 		\multicolumn{4}{l}{Bright sample ($i_{\rm AB}<21$)}\\
 		\hline
 		D4000$_{\rm cut\, w}$ & 1.54 (81.45\%) & 1.57 (89.86\%) & 1.57 (85.36\%)\\
 		D4000$_{\rm cut\, n}$ & 1.39 (76.36\%) [70.05\%] & 1.38 (88.99\%) [79.61\%] & 1.42 (84.10\%) [78.90\%]\\
  		\hline
 		\multicolumn{4}{l}{Full sample ($i_{\rm AB}<23$)}\\
 		\hline
  		D4000$_{\rm cut\, w}$ & 1.65 (75.67\%) & 1.58 (90.61\%) & 1.57 (87.20\%)\\
 		D4000$_{\rm cut\, n}$ & 1.53 (69.13\%) [69.03\%] & 1.38 (89.98\%) [84.75\%] & 1.40 (85.13\%) [81.33\%]\\
		\hline
 	\end{tabular}
\end{table*}

The PAUS direct D4000 measurement achieves the lowest percentage of correct classification due to the photometric noise, with at most 81\% of objects correctly classified (for the bright sample using the D4000$_{\rm w}$). This is also the instance where its effectiveness at classifying becomes closest to the VIPERS D4000 spectroscopic measurement (85\%). Using the full sample degrades the percentage of correct classification of the PAUS direct cut, given that there is a significant fraction of objects with $\rm SNR < 3$. This also translates in a flattening of the curve in \cref{fig:D4000_threshold}; in fact, for the D4000$_{\rm n}$ this curve is almost horizontal after the optimal D4000$_{\rm cut}$ value, which shows that varying D4000$_{\rm cut}$ barely has any effect on the reliability of the classification, and thus the correlation between galaxy type and D4000 value is weak.

PAUS CIGALE D4000 outperforms the VIPERS spectroscopic measurement in its respective optimal D4000$_{\rm cut}$ (at least 89\% of correctly classified objects). This is reasonable, given the model-driven bimodality/clustering that appears in the CIGALE D4000 reconstructed values, stemming from the SED templates (\cref{fig:D4000_hist}). Moreover, the PAUS CIGALE D4000$_{\rm cut}$ is the closest to spectroscopy in almost all cases (within $\pm0.02$, except for D4000$_{\rm cut \, n}$ for the bright sample). This shows that the PAUS CIGALE D4000 provides realistic D4000 values in the green valley between the red and blue populations, where the D4000 cut is placed. PAUS direct also yields values of D4000$_{\rm cut}$ close to the VIPERS results (within $\pm0.02$), but only for the bright sample (see \cref{tab:D4000_cut_results}); the mean SNR is too low in the full sample to provide a reliable galaxy classification.

It is worth noting that, for the case of D4000$_{\rm cut\, n}=1.55$ (values in brackets in \cref{tab:D4000_cut_results}), the difference in the percentage of correctly classified galaxies between the VIPERS D4000 spectroscopic measurement and the CIGALE D4000 reconstruction is significantly smaller (within $\sim$3\% maximum) than for our fiducial classification. This result is reasonable: if we use a D4000 cut already defined in the literature for spectroscopic D4000 measurements, the SED-fitting approach of CIGALE does not outperform the spectroscopic VIPERS D4000 nearly as much.

In order to check if these results were consistent with a different criterion for red/blue galaxy classification, we have recomputed the optimal D4000 threshold and its percentage of correctly classified objects, as in \cref{fig:D4000_threshold}, but using as a fiducial classification the spectroscopic VIPERS D4000$_{\rm cut}$ shown in \cref{tab:D4000_cut_results} for the bright sample. The results are largely the same (similar maximum percentage and optimal D4000 thresholds), and thus we can state that the performance of the D4000 estimations for galaxy classification is not largely dependent on the fiducial classification itself.

\subsection{D4000 versus stellar mass and SFR}\label{sec:D4000 versus stellar mass and SFR}

\begin{figure*}
 	 \includegraphics[width=\textwidth]{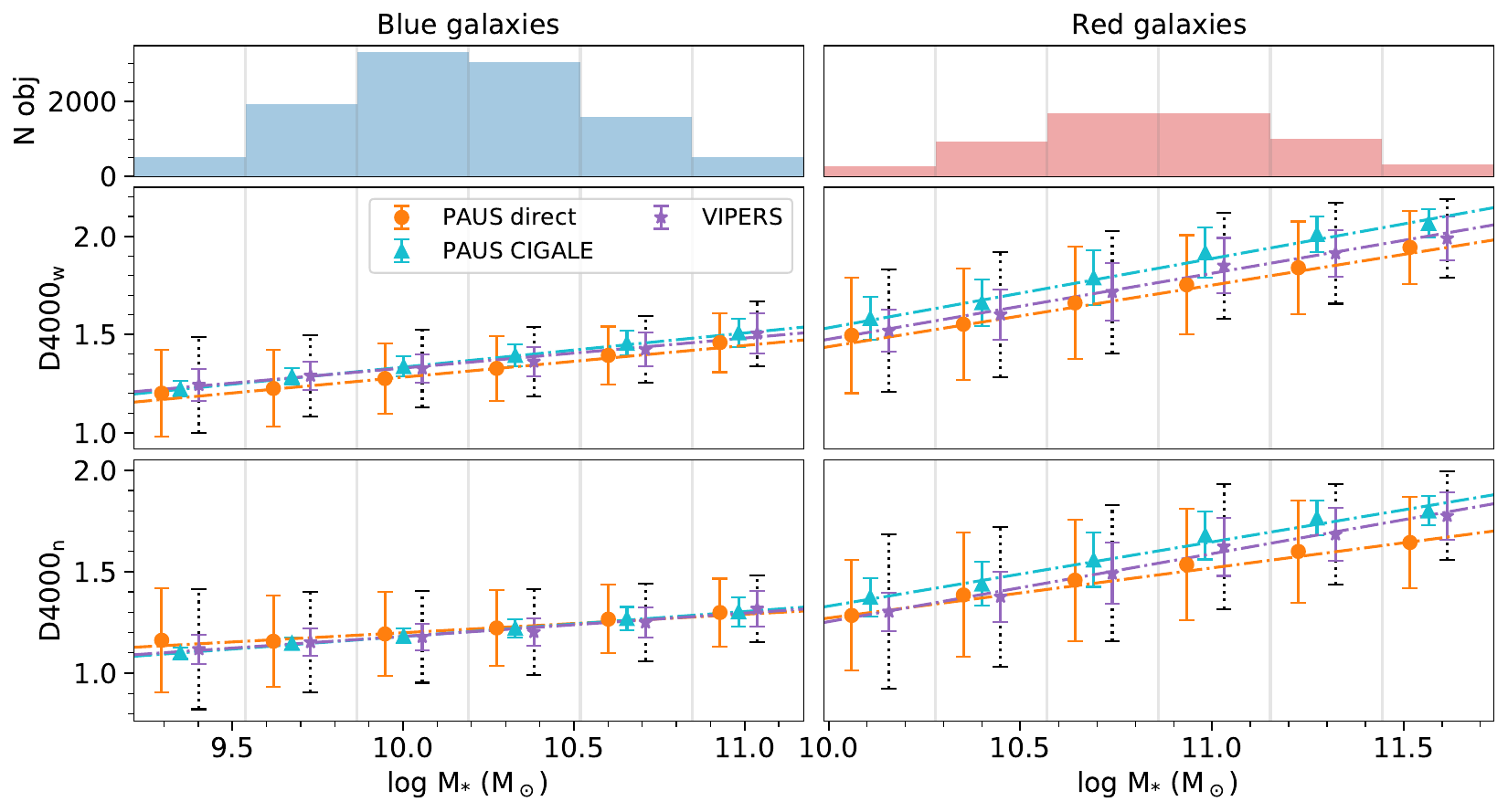}
     \caption{Median D4000 in uniform bins of log $\rm M_{*}$, for PAUS direct measurement (orange errorbars), PAUS CIGALE reconstruction (cyan) and VIPERS measurement (purple). The fits have been computed to the median values displayed in the plot. The errorbars represent the MAD of the D4000 of the sample. The black dotted errorbars in VIPERS show the MAD of the sample after adding to each object a Gaussian error with the same $\sigma$ as its respective PAUS direct measurement (as in \cref{fig:D4000_hist}). The errorbars for each measurement method have been offset in the x-axis for clarity, but in all cases the same binning has been used (as indicated by the grey vertical lines). Middle and lower rows are for D4000$_{\rm w}$ and D4000$_{\rm n}$ respectively, while left and right columns are for blue and red galaxies respectively, according to the classification of \citet{Siudek2018}. The upper row displays a histogram of the number of objects for the respective population.}
     \label{fig:D4000_mdn_vs_sm}
\end{figure*}

\begin{figure}
 	 \includegraphics[width=\columnwidth]{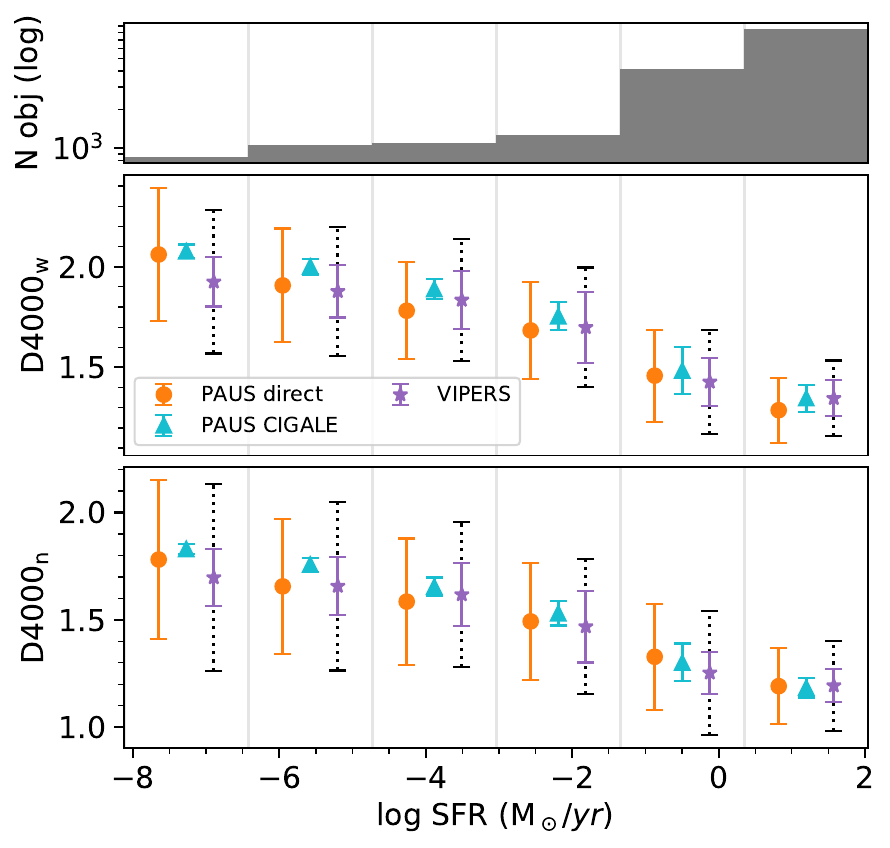}
     \caption{Median uniform bins of log SFR instead of log $\rm M_{*}$, following the colour code of \cref{fig:D4000_mdn_vs_sm}, but the red/blue classification. Upper panel shows a density histogram in logarithmic scale.}
     \label{fig:D4000_mdn_vs_sfr}
\end{figure}

The dependence of the D4000 versus stellar mass and SFR (\cref{fig:D4000_mdn_vs_sm} and \cref{fig:D4000_mdn_vs_sfr}, respectively) is evaluated for the full sample. For the D4000-mass relation, we separate the sample into red and blue red galaxies based on the fiducial classification (see \cref{sec:Galaxy classification}). The errorbars are the median absolute deviation (MAD) of the D4000 estimations. The MAD has been multiplied by its respective factor ($\sim$1.4826) so it is comparable to the standard deviation, and has not been divided by the square root of the number of objects per bin, so it accounts for the intrinsic variance of D4000 values inside a bin (therefore, it should remain constant for a number of objects large enough). We have used the median and the MAD to evaluate the dependence of the D4000 due to its insensitivity to outliers.

\begin{table}
\centering
	\caption{Slopes of all the linear fits displayed in \cref{fig:D4000_mdn_vs_sm}.}
 	\label{tab:D4000-mass_slopes}
 	\begin{tabular}{rlccc}
 		\hline
   	    \makecell{Galaxy \\ type} & \makecell{D4000 \\ definition} & \makecell{PAUS \\ direct} & \makecell{PAUS \\ CIGALE} & VIPERS\\
   	    \hline
   	    \multirow{2}{*}{Blue} & D4000$_{\rm w}$ & $0.16 \pm 0.01$ & $0.17 \pm 0.01$ & $0.15 \pm 0.01$\\
  	     & D4000$_{\rm n}$ & $0.09 \pm 0.01$ & $0.12 \pm 0.01$ & $0.11 \pm 0.01$\\
 		\hline
    	 \multirow{2}{*}{Red} & D4000$_{\rm w}$ & $0.31 \pm 0.01$ & $0.35 \pm 0.02$ & $0.34 \pm 0.02$\\
  	     & D4000$_{\rm n}$ & $0.25 \pm 0.01$ & $0.32 \pm 0.02$ & $0.33 \pm 0.02$\\
        \hline
 	\end{tabular}
\end{table}

In order to quantify the D4000-mass relation in \cref{fig:D4000_mdn_vs_sm} we have performed linear fits to the median values of the stellar mass bins; their slopes are specified in \cref{tab:D4000-mass_slopes}. The PAUS direct D4000-mass slope shows a good agreement with VIPERS for blue galaxies, with the slope values agreeing well within their error intervals derived from the fit. On the other hand, the slopes of the red population only agree for D4000$_{\rm w}$; for D4000$_{\rm n}$ PAUS direct significantly underestimates the slope. This is a consequence of the negative D4000-bias correlation shown in \cref{fig:scatter_bias}; for low (high) true D4000 values, the PAUS direct measurements overestimate (underestimate) the D4000. The MAD of the test dataset (VIPERS with PAUS direct $\sigma$) in \cref{fig:D4000_mdn_vs_sm} is also very similar to the MAD of PAUS direct D4000 for most stellar mass bins (especially for D4000$_{\rm w}$). This agreement reinforces the already discussed fact that the PAUS direct dispersion is fairly coherent with a Gaussian realisation of the error.

The PAUS CIGALE D4000 slopes agree with the VIPERS ones within their error intervals for all cases, with D4000$_{\rm w}$ and D4000$_{\rm n}$ performing equally well (\cref{tab:D4000-mass_slopes}). However, the CIGALE D4000 reconstruction shows in the high-mass end of the red sample and the low-mass end of the blue sample a smaller MAD than the VIPERS D4000 (by a factor of $\sim$3 and $\sim$2, respectively). This underestimation is a clear sign that, despite correctly determining the average spectral features of a given population, the SED templates may not fully reproduce the actual diversity of the sample in the most "extreme" regimes: the bluest and least massive star-forming galaxies, and the most massive and quenched red galaxies.

We can compare the slopes from \cref{tab:D4000-mass_slopes} with previous results from the literature. In \citet{Haines2017}, the D4000$_{\rm n}$-mass relation is derived for red and blue galaxies separated by a D4000$_{\rm cut \, n}=1.55$, using the same VIPERS data release \citep[][]{Scodeggio2018}. They reported a blue D4000$_{\rm n}$-mass slope between 0.12 and 0.14, invariant with redshift; a result reasonably close to the slope of 0.11 we report in \cref{tab:D4000-mass_slopes}. Regarding the red slope, \citet{Haines2017} derived a value of $\sim$0.23 at $0.5<z<1.0$, a slope significantly less steep that the value of 0.33 we find. Another work that analyses the D4000$_{\rm n}$-mass relation for the red sequence is \citet{Siudek2017}, where an evolving $U-V$ colour cut is used to select red galaxies. Here, a slope of $0.164 \pm 0.031$ is found for the same VIPERS data release at $ 0.40 < z < 1.00$; a value in even larger disagreement with our results.

The reason behind these discrepancies in the red slope lies in the different red/blue classification methods used in each work (and to a lesser extent,  the differences in sample selection). If a different classification method results in a given percentage of "contamination" compared to our classification (i.e., our blue galaxies being identified as red in another work, and vice-versa), this contamination will significantly affect the red sequence, as only 35\% of our sample is composed of red galaxies, according to our fiducial classification (i.e, see the histograms in \cref{fig:D4000_mdn_vs_sm}). In comparison, the blue cloud will result mostly unaffected by contamination; hence the reasonable agreement between our D4000$_{\rm n}$-mass blue slope and the results reported in \citet{Haines2017}. The physical mechanism behind the increasing D4000$_{\rm n}$-mass trend reported both here and in \citet{Haines2017,Siudek2017}, is the downsizing scenario \citep{Cowie1996}, widely tested in the literature: more massive galaxies are older, and thus with higher D4000 values.

The D4000-SFR relation (see \cref{fig:D4000_mdn_vs_sm}) is in good agreement with spectroscopy for both the PAUS direct measurement and the PAUS CIGALE reconstruction, with differences in median D4000 smaller than 0.1 for most cases (only a significant overestimation of $\sim$0.2 appears in the lowest-SFR bin). The MAD of the test dataset (VIPERS D4000 with PAUS direct noise) is again similar to that of the PAUS direct measurement. The only significant caveat that arises is the underestimation of the MAD for the PAUS CIGALE D4000 reconstruction when compared to the VIPERS D4000 measurement, especially for the very low and highest SFR bins. This is consistent with our findings in the D4000-mass relation, and again shows that the SED templates do not properly reproduce the true galaxy diversity in the most quenched and star-forming regimes.

With the dependency of D4000 with galaxy properties shown in this subsection, and the similar trends found in the literature (with discrepancies stemming from different red/blue selection biases), we see that narrow-bands surveys show a great potential to break the incompleteness issue of flux-limited and target-based spectroscopy, as the direct photometric measurement of D4000 retrieves average trends compatible with spectroscopy. Moreover, the study of environmental dependence with statistical significance so far not achievable by spectroscopic surveys \citep[][]{Siudek2022} could also be performed with narrow-band data. The synergy of wide coverage with accurate measurements of spectral features may open a new insight in the nature vs nurture dilemma \citep[e.g.,][]{Peng2010, DeLucia2012}.

\section{Conclusions}\label{sec:Conclusions}

We have carried out the first in-depth study of the direct photometric measurement of the D4000 with narrow-band photometry, using the observational data of PAUS cross-matched with VIPERS \citep{Scodeggio2018} in the CFHTLS W1 field. The results of the photometric D4000 measurement have been compared to its spectroscopic counterpart, as well as the D4000 reconstructed from the SED-fitting code CIGALE (with either CFHTLS broad bands or PAUS NBs). Both the original definition of the spectral break from \citet{Bruzual1983}: D4000$_{\rm w}$, and the narrower definition from \citet{Balogh1999}: D4000$_{\rm n}$, have been considered.

First, we have developed a general D4000 estimator for narrow-band photometry, and we have tested it by measuring the D4000 in a synthetic PAUS catalogue, where we find that over 95\% (92\%) of objects have an absolute bias below 5\% in the PAUS D4000$_{\rm w}$ (D4000$_{\rm n}$) measurement. We also find similar bias trends to the preliminary work of \citet{Stothert2018}, but with a significant improvement (the average bias for D4000$_{\rm w}$ is reduced by $\sim$2). With real PAUS narrow-band observations, only 85.01\% (D4000$_{\rm w}$) and 65.87\% (D4000$_{\rm n}$) have $\rm SNR > 3$. However, an $i_{\rm AB}<21$ cut results in a bright sample where virtually all of the objects have $\rm SNR > 3$. Hence, the PAUS catalogue achieves "D4000-completeness" two magnitudes below the actual magnitude limit. The D4000 reconstructions from CIGALE present much higher SNR, but their errors are underestimated by $>$50\%. All estimation methods show similar bias trends: an absolute median bias of few \% and a negative bias-VIPERS D4000 correlation. We also have compared D4000 distributions, and find that the PAUS direct D4000 follows closely the VIPERS D4000 with photometric Gaussian noise, and that the CIGALE reconstruction is far more realistic if PAUS NBs are used instead of CFHTLS broad bands.

We have evaluated the D4000-redshift relation, and find that the PAUS direct measurements are fully compatible with VIPERS D4000, but the PAUS CIGALE D4000 overestimates the slope by a factor of $\sim$2. Moreover, we have also examined the potential of these D4000 estimations for galaxy classification by comparing a given D4000$_{\rm cut}$ value to a fiducial classification \citep{Siudek2018}. The D4000 reconstruction from PAUS CIGALE outperforms even VIPERS spectroscopy (>90\% of correctly classified objects vs >85\%), while the PAUS direct measurements perform poorly due to the larger scatter (81\% of correctly classified objects at best). Finally, we have evaluated the D4000-stellar mass and D4000-SFR relation, where we find a generally good agreement between estimation methods. For the D4000-mass relation, we have separated the sample in red/blue galaxies, using the same fiducial classification, and find a blue slope reasonably close to previous results \citep[][]{Haines2017}. The red slope seems to be steeper than in other works \citep[e.g.,][]{Haines2017, Siudek2017}, due to the different red/blue classifications employed.

In conclusion, this work shows that narrow-band photometry with PAUS-like spectral resolution ($R\sim65$) can be used to directly measure the D4000 in a model-independent way; all objects two magnitudes below the magnitude limit can be individually measured with $\rm SNR > 3$. For fainter magnitudes, stacking low-SNR measurements yields results consistent with spectroscopy. The D4000 reconstruction via SED fitting greatly benefits from the use of narrow-band photometry, providing a far more realistic distribution. However, it presents some artefacts such as noise underestimation, artificial bimodality or underestimation of sample variance. 

Therefore, these results open up several possibilities for narrow-band photometry that may not be available to conventional spectroscopy, such as the study of the radial dependence at different apertures (stacking several objects if necessary). More conventional studies can also be carried out, e.g., exploring the star formation history by constraining their stellar ages and redshift of formation. The wealth of narrow-band photometric data from PAUS may also be used to further study the relationship between morphology and D4000/galaxy quenching \citep[e.g., see][and the references therein]{Kim2018}, or to analyse the role of the environment in these processes, and thus disentangle the influence of nature vs nurture in galaxy evolution \citep[e.g.,][]{Peng2010, DeLucia2012}.

\section*{Acknowledgements}

The PAU Survey is partially supported by MINECO under grants CSD2007-00060, AYA2015-71825,  ESP2017-89838, PGC2018-094773, PGC2018-102021, SEV-2016-0588, SEV-2016-0597, MDM-2015-0509 and Juan de la Cierva fellowship and LACEGAL and EWC Marie Sklodowska-Curie grant No 734374 and no.776247 with ERDF funds from the EU  Horizon 2020 Programme, some of which include ERDF funds from the European Union. IEEC and IFAE are partially funded by the CERCA and Beatriu de Pinos program of the Generalitat de Catalunya. Funding for PAUS has also been provided by Durham University (via the ERC StG DEGAS-259586), ETH Zurich, Leiden University (via ERC StG ADULT-279396 and Netherlands Organisation for Scientific Research (NWO) Vici grant 639.043.512), University College London and from the European Union's Horizon 2020 research and innovation programme under the grant agreement No 776247 EWC. The PAU data center is hosted by the Port d'Informaci\'o Cient\'ifica (PIC), maintained through a collaboration of CIEMAT and IFAE, with additional support from Universitat Aut\`onoma de Barcelona and ERDF. We acknowledge the PIC services department team for their support and fruitful discussions. P.R. and Z.C. are supported by  National Science Foundation of China (grant No. 12073014). The results published have been funded by the European Union's  Horizon 2020 research and innovation programme under the Maria Skłodowska-Curie (grant agreement No 754510), the National Science Centre of Poland (grant UMO-2016/23/N/ST9/02963) and by the Spanish Ministry of Science and Innovation through Juan de la Cierva-formacion program (reference FJC2018-038792-I). AHW is supported by an European Research Council Consolidator Grant (No. 770935).

We also thank the anonymous referee for their their effort reviewing the paper and their insightful comments.

\section*{Data Availability}

The data underlying this article will be shared on reasonable request to the corresponding author.



\bibliographystyle{mnras}
\bibliography{main} 




\appendix

\section{Redshift comparison}\label{sec:Redshift comparison}

\begin{figure}
 	\includegraphics[width=\columnwidth]{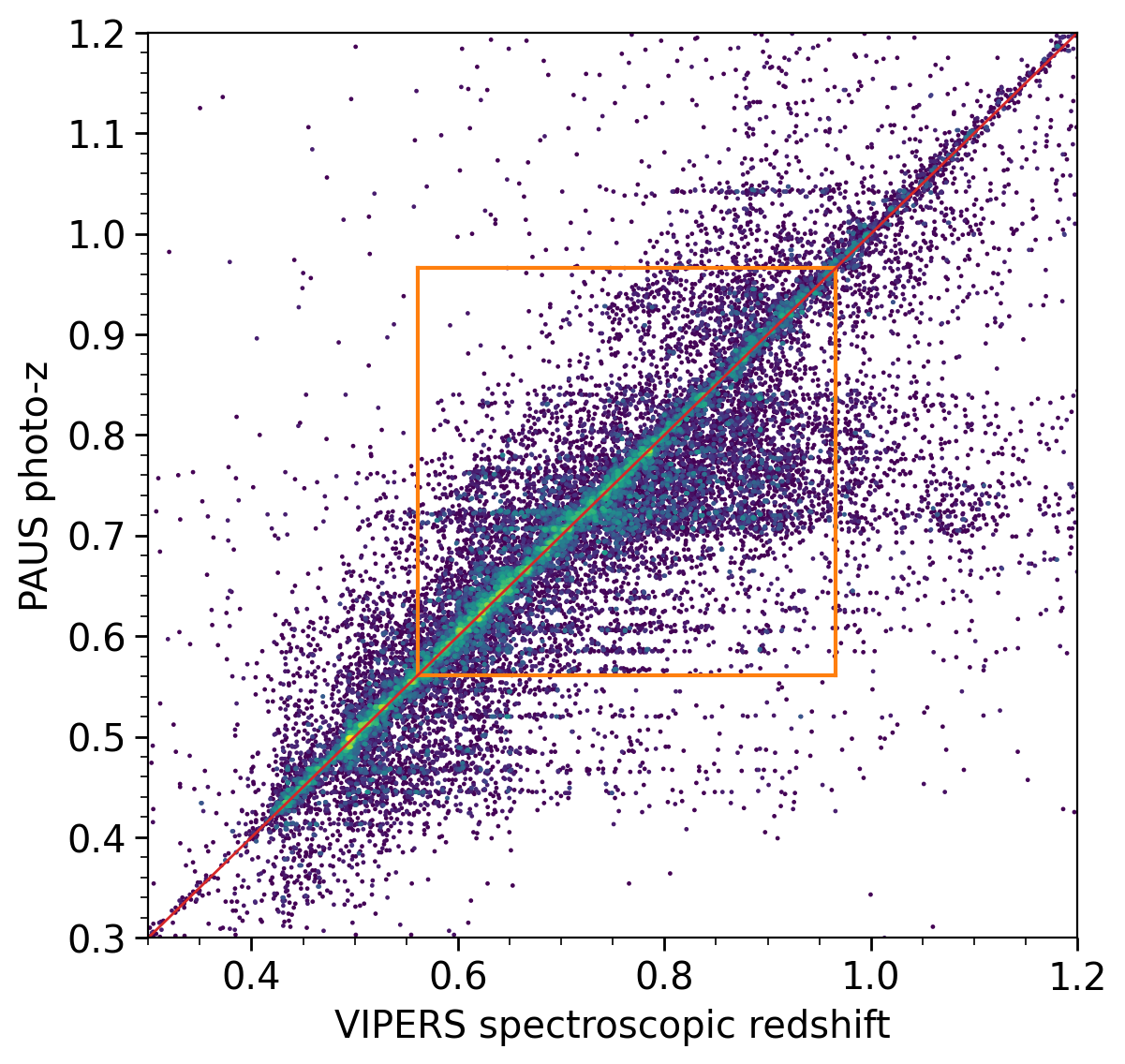}
     \caption{PAUS photo-z versus VIPERS spec-z, for the cross-match of 28.788 galaxies with accurate spec-z measurements. Points are coloured according to the logarithm of the point density (with lighter areas having more objects). The red line represents the 1:1 equivalence between redshift measurements. The redshift cuts introduced for the D4000 measurement, both in PAUS photo-z and VIPERS spec-z, are represented by the orange box.}
     \label{fig:z_comparison}
\end{figure}

A comparison of the distribution of PAUS photo-z and VIPERS spec-z for the W1 cross-match is shown in \cref{fig:z_comparison}. The orange box in the middle of the plot represents the $0.562<z<0.967$ cut in both redshift estimations; hence all objects inside the box are the selected sample for D4000 measurements. Most of the PAUS photo-z follows closely along the red diagonal line representing the 1:1 equivalence. It is worth noting that the determination of PAUS photo-z in the wide fields is still a work in progress, and thus some artefacts that can be appreciated in \cref{fig:z_comparison} (horizontal stripes) may disappear in future data releases.

\begin{figure}
 	\includegraphics[width=\columnwidth]{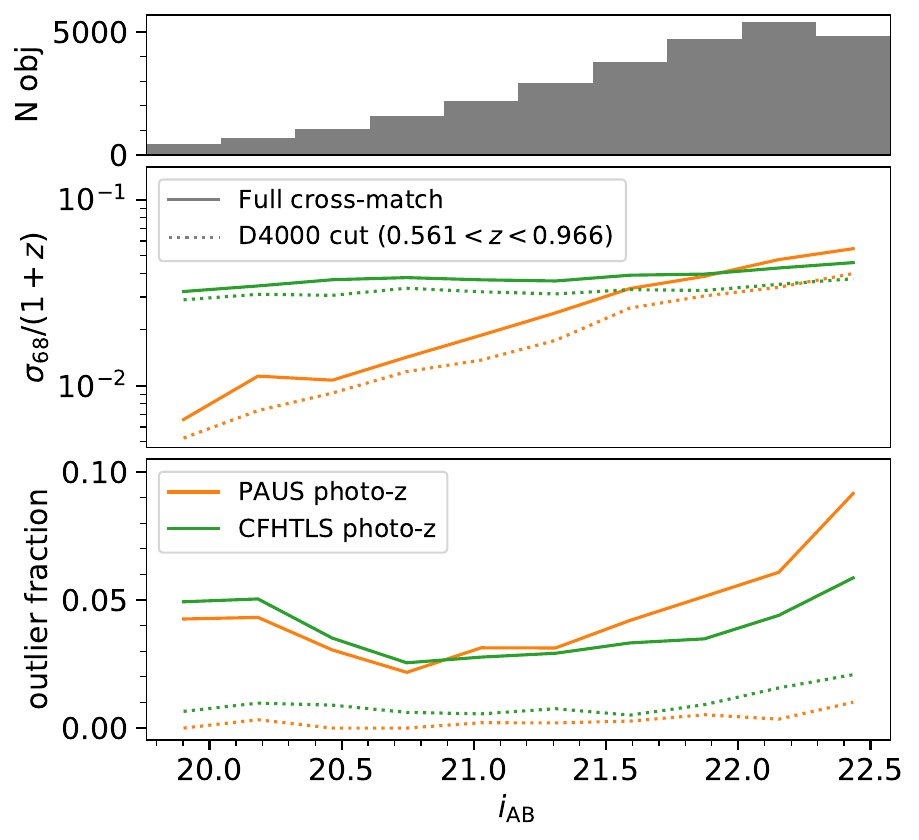}
     \caption{$\sigma_{68}/(1+z)$ (middle panel) and outlier fraction (lower panel) for the PAUS photo-z (orange lines) and CFHTLS photo-z (green lines), in magnitude bins. Solid line represents the full cross-match, dotted line the redshift cut for D4000 measurement. Upper panel displays the magnitude histogram.}
     \label{fig:z_sigma_outlier}
\end{figure}

We can further evaluate the performance of PAUS photo-z with its dispersion, $\sigma_{68}/(1+z)$. If we first define $\Delta z$ as

\begin{equation}\label{eq:delta_z}
    \Delta z = \frac{z_{\rm PAUS}-z_{\rm spec}}{1+z_{\rm spec}},
\end{equation}

we can define $\sigma_{68}$ as

\begin{equation}\label{eq:sigma68}
    \sigma_{68} = \frac{P_{\rm 84.1\%}(\Delta z) - P_{\rm 15.9\%}(\Delta z)}{2},
\end{equation}

where $P_{i \%}$ is the \textit{i}th percentile of the distribution. This $\sigma_{68}$ is equivalent to $\sigma$ if $\Delta z$ had a Gaussian distribution, while being much less sensitive to outliers. For this work, outliers have been defined as all galaxies which have $|\Delta z| > 0.15$, following \citet{Soo2021}. In \cref{fig:z_sigma_outlier}, the average $\sigma_{68}/(1+z)$ value, as well as the outlier fraction, is shown for both the PAUS photo-z and the broad-band photo-z from the CFHTLS reference catalogue \citep{Ilbert2006, Coupon2009}. For the D4000 redshift cut, the outlier fraction for PAUS is kept below 2\%, and PAUS photo-z shows a smaller scatter than the CFHTLS catalogue for the whole magnitude range ($\langle\sigma_{68}/(1+z)\rangle=0.019$ for PAUS, $\langle\sigma_{68}/(1+z)\rangle=0.032$ for CFHTLS).

For the full redshift range, however, the outlier fraction of PAUS photo-z seems significantly larger than CFHTLS at $i_{\rm AB} > 21.5$; this is mostly due to the $z<1.2$ limit in the photo-z run, and does not affect the sample selected for this work. The $\sigma_{68}$ of PAUS photo-z also is slightly larger than its CFHTLS counterpart at the faint magnitude end ($i_{\rm AB} > 22$); given that both PAUS NBs and CFHTLS broad-band data are used in the PAUS photo-z calculation \citep{Eriksen2019}, ideally the PAUS photo-z should never perform worse than CFHTLS alone. Hence, this higher PAUS $\sigma_{68}$ is an issue of the NB weighting relative to broad-band data, which is currently being worked on. Also, the worse performance of PAUS photo-z shown here when compared to earlier works in the COSMOS field \citep[$\langle\sigma_{68}/(1+z)\rangle\simeq0.0037$ for a 50\% quality cut;][]{Eriksen2019, Alarcon2021} is mostly due to the smaller number of exposures and higher mean redshift in the wide fields.

\bsp	
\label{lastpage}
\end{document}